\documentclass[12pt]{article}
\usepackage{jheppub}
\title{Two Dimensional Renormalization Group Flows in
Next to Leading Order}

\author[]{Rubik Poghossian}

\affiliation[]{Yerevan Physics Institute\\
Alikhanian Br. 2, 0036 Yerevan, Armenia
\\
e-mail: poghos@yerphi.am}

\abstract{Zamolodchikov's famous analysis of the RG trajectory connecting
successive minimal CFT models $M_p$ and $M_{p-1}$ for $p\gg 1$, is
improved by including second order in coupling constant
corrections. This allows to compute IR quantities with next to
leading order accuracy of the $1/p$ expansion. We compute
in particular, the beta function and the anomalous dimensions
for certain classes of fields. As a result we are able to identify with a
greater accuracy the IR limit of these fields with certain linear
combination of the IR theory $M_{p-1}$.
We discuss the relation of these results with Gaiotto's recent RG domain
wall proposal.
}
\newcommand{\ie}{{\it i.e.\ }}
\def\bea{\begin{eqnarray}}
\def\eea{\end{eqnarray}}
\def\a{\alpha}
\def\b{\beta}

\def\G{\Gamma}
\def\d{\delta}
\def\D{\Delta}
\def\e{\epsilon}

\def\l{\lambda}

\begin{document}
\maketitle

\section*{Introduction}
In his famous paper \cite{Zamolodchikov:1987} A.~Zamolodchikov has investigated
Renormalization Group (RG) flow from the minimal model
$M_p$ to the $M_{p-1}$ for large $p\gg1$ caused
by the relevant operator $\phi_{1,3}$. Two main circumstances
made it possible to investigate this RG flow using a single coupling
constant perturbation theory. First, the conformal
dimension of the field $\phi_{1,3}$, $\D_{1,3}=1-\frac{2}{p+1}\equiv 1-\e$
(see Appendix \ref{A}) is
nearly marginal when $p\gg 1$ and second, the Operator
Product Expansion (OPE) of this field with itself produces no relevant field
besides the initial one and the unit operator. The method of
A.~Zamolodchikov not only allowed to identify the IR theory with $M_{p-1}$,
but also provided detailed description how several
classes of local fields behave along the RG trajectory. The analogous RG flow
for the $N=1$ super minimal models has been investigated in \cite{Poghossian:1988}.

The main purpose of this paper is a sharpening of Zamolodchikov's analysis,
by the inclusion of {\it second order} perturbative corrections.
It is interesting to note that in all cases we have investigated,
the rotation matrix (in the space of fields), that diagonalizes the
matrix of anomalous dimensions, does not receive $1/p$
or $1/p^2$ corrections. So an interesting question arises, if any higher
order corrections appear at all.

As intermediate results, in this paper we have found several four-point
correlation functions in large $p$ limit (see formulae (\ref{4pfinfp})).

The initial motivation to carry out these computations came
from the recent approach to this RG flow by D.~Gaiotto \cite{Gaiotto:2012RGDW}. Using
Goddard-Kent-Olive construction, Gaiotto has constructed a non-trivial
conformal interface between two successive minimal models and made a
striking conjecture, that this interface is the exact RG domain wall which
encodes the map between the UV and IR fields. Gaiotto's conjecture
survives a strong test: it is fully compatible with the first order
parturbative calculations of the mixing amplitudes performed by Zamolodchikov.

In this paper we show that this mixing coefficients computed with the
help of the perturbation theory up to the second order, unlike those obtained
from the Gaiotto's conjecture,
do not receive any corrections up to the order $1/p^2$. Nevertheless, this
discrepancy might be attributed to the renormalization scheme
which is adopted here following Zamolodchikov. Presently the author
of this paper does not have any clue how to take into account possible
dependencies on the renormalization schemes in order to be able to
make any conclusive statement about Gaiotto's conjecture beyond the
leading order.

The paper is organized as follows.

In Section \ref{PTSO}, we develope some technical tools, necessary to carry
out second order in coupling constant calculations.

In Section \ref{bf} the $\b$-function and Zamolodchikov's $c$-function
\cite{Zamolodchikov:1986} are
computed with next to leading order accuracy. The critical value of the
renormalized coupling constant, the slope of the $\b$-function as well as
the $c$-function at the critical point are calculated. The results of these computations
confirm that also the second order contributions perfectly match with
the Zamolodchikoved's conclusion that the IR fixed point corresponds to the
CFT $M_{p-1}$ and that the UV field $\phi_{1,3}$ flows to the field
$\phi_{3,1}$ of the IR theory.

Section \ref{FRUIM} is devoted to the renormalization of several
series of local fields and to the calculation of their anomalous dimensions.
Thus:\\
in Section \ref{PFnn} we investigate the renormalization of the fields $\phi_{n,n}$.\\
In Section \ref{Pnp1nm1} the renormalization of the fields $\phi_{n,n+1}$ and
$\phi_{n,n-1}$ is discussed and the matrix of anomalous dimensions
is found. At the fixed point the matrix of anomalous dimensions
is diagonalized and its eigenvalues are calculated.\\
In Section \ref{Pnp2nm2} the same steps are performed for the fields
$\phi_{n,n+2}$, $\partial \bar{\partial}\phi_{n,n}$ and $\phi_{n,n+2}$.

In all cases the predictions of Zamolodchikov successfully withstand
the next to leading order test.

In Appendix \ref{A} some basic facts about the minimal models of 2d CFT
are reviewed. The Appendices \ref{B} and \ref{C} are devoted to computation
of the integrals used in the main text. The Appendix \ref{D} comments how to
calculate the large $p$ limit of those four point correlation
functions used in the main text.

\section{Perturbation theory in second order}
\label{PTSO}
Suppose the (Euclidean) action density is given by
\bea
{\cal H}(x)={\cal H}_0(x)+\l \phi(x)
\label{action}
\eea
with ${\cal H}_0$ being the UV CFT action density, $\phi$ a relevant
local spinless field and $\l$ the coupling constant.
Then for a two-point function up to second order we'll have
\bea
\langle \phi_1(y_1)\phi_2(y_2)\rangle_\l &=&\langle \phi_1(y_1)\phi_2(y_2)\rangle_0
-\l \int\langle \phi_1(y_1)\phi_2(y_2)\phi(x)\rangle_0 d^2x\nonumber\\
&+&\frac{\l^2}{2}\int\langle \phi_1(y_1)\phi_2(y_2)\phi(x_1)
\phi(x_2)\rangle_0 d^2x_1d^2x_2+O(\l^3)
\label{2pfunclam}
\eea
In this paper we consider a theory, whose UV limit is given by the minimal
CFT model $M_p$ with $p\gg 1$ and the perturbing field is $\phi\equiv \phi_{1,3}$.
Leading order corrections in this theory has been
investigated by A.~Zamolodchikov \cite{Zamolodchikov:1987}.
Second order computations are more complicated. Indeed, not only the knowledge of
four point correlation functions which in general are quite non-trivial in a CFT \cite{BPZ:1984}, but also their integrals over two insertion points is required.
Fortunately, as we demonstrate below, the conformal invariance allows to perform
integration over one of the insertion points explicitly.
First let us notice that translational and scale invariance can be exploited to
locate the points $y_1$, $y_2$ at $y_1=1$ and $y_2=0$ without loss of generality:
\bea
&&\int\langle \phi_1(y_1)\phi_2(y_2)\phi(x_1)
\phi(x_2)\rangle_0 d^2x_1d^2x_2\nonumber\\
&&=(y_{12}\bar{y}_{12})^{2-2\D-\D_1-\D_2}
\int\langle \phi_1(1)\phi_2(0)\phi(x_1)
\phi(x_2)\rangle_0 d^2x_1d^2x_2
\label{2ndorderint}
\eea
(here and below I frequently use the shorthand notation
$x_{12}=x_1-x_2$, $y_{12}=y_1-y_2$ et.al.).
Any four-point function of primary fields in a CFT essentially
depends only on the cross ratio
$x=\frac{x_{12}x_{34}}{x_{14}x_{32}}$ of the insertion
points and its conjugate \cite{BPZ:1984}
\bea
\label{gen4p}
\langle \phi_1(x_1)\phi_2(x_2)\phi_3(x_3)
\phi_4(x_4)\rangle
&=& (x_{14}\bar{x}_{14})^{-2 \D_1}
(x_{24}\bar{x}_{24})^{\D_1+\D_3-\D_2-\D_4}\\
&\times& (x_{34}\bar{x}_{34})^{\D_1+\D_2-\D_3-\D_4}
(x_{23}\bar{x}_{23})^{\D_4-\D_1-\D_2-\D_3}G(x,\bar{x}), \nonumber
\eea
where it is assumed that the fields are spin-less (i.e. $\D_i=\bar{\D}_i$).
Specifying the insertion points as $x_1=x$, $x_2=0$, $x_3=1$ and $x_4=\infty$
we get
\bea
G(x,\bar{x})=\lim_{x_4\rightarrow\infty}(x_4\bar{x}_4)^{2\D_4}\langle \phi_1(x)\phi_2(0)\phi_3(1)
\phi_4(x_4)\rangle\equiv \langle \phi_1(x)\phi_2(0)\phi_3(1)
\phi_4(\infty)\rangle
\label{G}
\eea
Alternatively specifying $x_1=1/x$, $x_2=\infty$, $x_3=1$ and $x_4=0$
and comparing with (\ref{G}) we get the identity
\bea
\langle \phi_1(x)\phi_2(0)\phi_3(1)
\phi_4(\infty)\rangle=(x\bar{x})^{-2
\D_1}\langle \phi_1(1/x)\phi_4(0)\phi_3(1)
\phi_2(\infty)\rangle
\label{Ginf}
\eea
which is useful when investigating the correlation functions at large $x$.
After application of (\ref{gen4p}), (\ref{G}) to the four-point function
$\langle \phi(x_1)\phi_2(0)\phi_1(1)\phi(x_2)\rangle_0$ and introduction of the
new integration variables
\bea
\frac{x_1(1-x_2)}{x_{12}}\rightarrow x_1; \qquad\qquad 1-x_2\rightarrow x_2.
\nonumber
\eea
two integrations on the r.h.s. of eq. (\ref{2ndorderint}) become partly
disentangled
\bea
\int\langle \phi_1(1)\phi_2(0)\phi(x_1)
\phi(x_2)\rangle_0 d^2x_1d^2x_2=\int I(x_1) \langle \phi(x_1)\phi_2(0)\phi_1(1)\phi(\infty)
\rangle_0d^2x_1
\label{2ndorderintreduced}
\eea
where
\bea
I(x)=\int (y\bar{y})^{a-1}((1-y)(1-\bar{y}))^{b-1}
((x-y)(\bar{x}-\bar{y}))^{c}d^2y,
\label{I(x)}
\eea
with parameters
\bea
a=\epsilon_{12}+2\epsilon;\quad b=\epsilon_{21}+2\epsilon ;\quad c=-2\epsilon
\label{abc}
\eea
where $\epsilon=1-\D$, $\epsilon_{1,2}=1-\D_{1,2}$ are
the complementary dimensions.
Fortunately the integral (\ref{I(x)}) can be expressed in terms of hyper-geometric
functions (see appendix \ref{B}):
\bea
\label{Iabc}
I(x)&=&\frac{\pi \gamma(b)\gamma(a+c)}{\gamma(a+b+c)}|F(1-a-b-c,-c,1-a-c,x)|^2\\
&+&\frac{\pi \gamma(1+c)\gamma(a)}{\gamma(1+a+c)}|x^{a+c}F(a,1-b,1+a+c,x)|^2\nonumber\\
&=&\frac{\pi \gamma(a)\gamma(b+c)}{\gamma(a+b+c)}|F(1-a-b-c,-c,1-b-c,1-x)|^2\nonumber\\
&+&\frac{\pi \gamma(1+c)\gamma(b)}{\gamma(1+b+c)}|(1-x)^{b+c}F(b,1-a,1+b+c,1-x)|^2\nonumber\\
&=&\frac{\pi \gamma(a)\gamma(b)}{\gamma(a+b)}|x^cF(a,-c,a+b,1/x)|^2\nonumber\\
&+&\frac{\pi\gamma(1+c)\gamma(a+b-1)}{\gamma(a+b+c)}|x^{a+b+c-1}
F(1-a-b-c,1-b,2-a-b,1/x)|^2\nonumber
\eea
where $\gamma(x)=\Gamma(x)/\Gamma(1-x)$ and $F(a,b,c,x)$ is
the Gaussian hypergeometric function.
Above three expressions for $I(x)$ are convenient when exploring
the regions $x\sim 0$, $x\sim 1$ and $x\sim \infty $ respectively.
Note also that these expressions make explicit the single-valuedness
of $I(x)$. Specifying the choice of parameters to (\ref{abc}) and applying the identity
\[
F(a,b,c,x)=(1-z)^{c-a-b}F(c-a,c-b,c,x)
\]
to the second term of the second equality, the eqs. (\ref{Iabc}) can be rewritten as
\bea
I(x)&=&\frac{\pi \gamma(2 \e+\e_{21})\gamma(\e_{12})}{\gamma(2\e)}
|F(1-2\e,2\e,1+\e_{21},x)|^2\nonumber\\
&+&\frac{\pi\gamma(2\e+\e_{12})\gamma(\e_{21})}{\gamma(2\e)}
|(x/(1-x))^{\e_{12}}F(2\e,1-2\e,1+\e_{12},x)|^2\nonumber\\
&=&\frac{\pi \gamma(2\e+\e_{12})\gamma(\e_{21})}{\gamma(2\e)}
|F(1-2\e,2\e,1+\e_{12},1-x)|^2\nonumber\\
&+&\frac{\pi\gamma(2\e+\e_{21})\gamma(\e_{12})}{\gamma(2\e)}
|(x/(1-x))^{\e_{12}}F(2\e,1-2\e,1+\e_{21},1-x)|^2\nonumber\\
&=&\frac{\pi \gamma(2\e+\e_{12})\gamma(2\e+\e_{21})}{\gamma(4\e)}
|x^{-2\e}F(2\e+\e_{12},2\e,4\e,1/x)|^2\nonumber\\
&+&\frac{\pi\gamma(4\e-1)}{\gamma^2(2\e)}|x^{2\e-1}F(1-2\e,1-2\e+\e_{12},2-4\e,1/x)|^2
\label{Ieps}
\eea
It is worth noting that in the case when $\e_{12}\equiv\e_1-\e_2=0$ only
the third expression is manifestly nonsingular, the first two
expressions require a subtle limiting procedure. Thus for
this case it is better to employ the third expression:
\bea
I(x)&=&\frac{\pi \gamma^2(2\e)}{\gamma(4\e)}
|x^{-2\e}F(2\e,2\e,4\e,1/x)|^2\nonumber\\
&+&\frac{\pi\gamma(4\e-1)}{\gamma^2(2\e)}|x^{2\e-1}F(1-2\e,1-2\e,2-4\e,1/x)|^2
\label{I0eps12}
\eea
Let us investigate the behaviour of (\ref{I0eps12}) at $x\sim 1$.
Using standard formulae for the analytic continuation of the
hypergeometric function with parameters satisfying the condition
$a+b-c \in \mathbb{Z}$ (see e.g. \cite{BatErd1}) one can get convinced that
\bea
I(x)&\approx &\pi  (x+\bar{x}-4)
+\pi  \left(1+2 \e (2 \epsilon -1) (x+\bar{x}-2)\right)\nonumber\\
&\times&\left(2-\log |x-1|^2-2 \pi  \cot (2 \pi  \epsilon )
-4 \psi (2 \epsilon )-4 \gamma \right)
\label{I0e121lim}
\eea
where $\gamma =0.577216\cdots$ is the Euler constant and the omitted terms are at most of order $|x-1|^2\log |x-1|$ in $x\rightarrow 1$ limit.
There is no need to investigate the limit $x\rightarrow0$ separately
since the obvious symmetry of $I(x)$ with respect to $x\leftrightarrow1-x$
at $x\sim0$ immediately ensures
\bea
I(x)&\approx &-\pi  (2+x+\bar{x})
+\pi  \left(1+2 \e (1-2 \e ) (x+\bar{x})\right)\nonumber\\
&\times&\left(2-\log |x|^2-2 \pi  \cot (2 \pi  \epsilon )
-4 \psi (2 \epsilon )-4 \gamma \right)
\label{I0e120lim}
\eea


\section{$\b$-function}
\label{bf}
\begin{figure}[htb]
\includegraphics[scale=0.3]{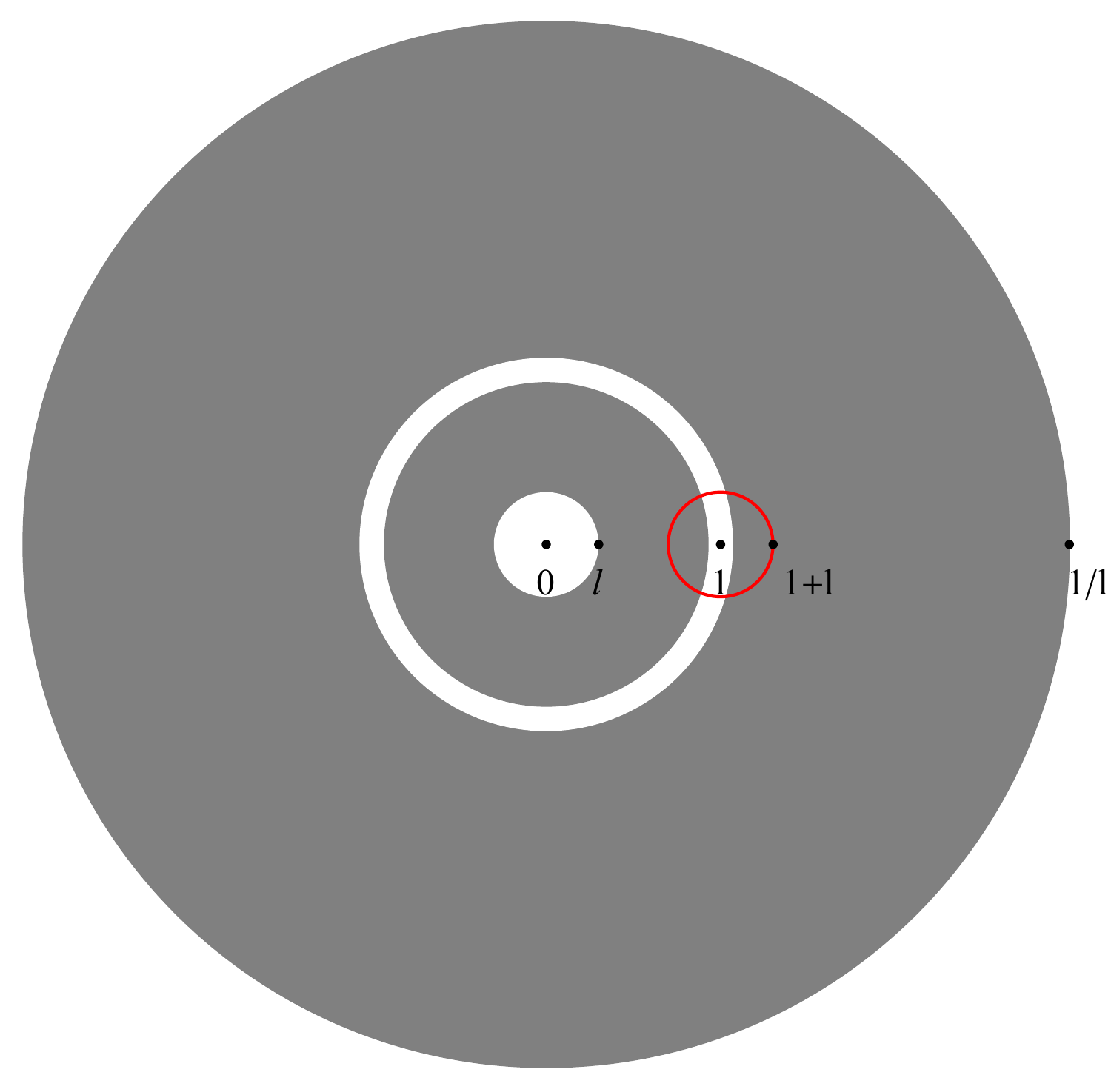}
\centering
\caption{$\Omega_{l,l_0}$ is the gray region}
\label{Fig1}
\end{figure}

In this section we calculate the $\b$-function up to $1/p^4\sim \e^4$ corrections
for the small values of the (renormalized) coupling constant (of order
$\e$ or smaller). As it will become quite clear  later for this purpose  one should
evaluate the integral (\ref{2ndorderintreduced}) in the special case $\phi_1=\phi_2=\phi$
and $I(y)$ given by (\ref{I0eps12}) with the accuracy $\sim 1/\e$.
Our strategy will be as follows: separate in the integration region the discs
$D_{l,0}=\{x\in \mathbb{C}\,|\,\,|x|<l\}$, $D_{l,1}=\{x\in \mathbb{C}\,|\,\,|x-1|<l\}$
and $D_{l,\infty}=\{x\in \mathbb{C}\,|\,\,|x|>1/l\}$ where $l$ is an intermediate
length scale such that $0<l_0 \ll \exp(-1/\e)\ll l\ll 1$ and $l_0$ is the
ultraviolet scale. For the integral outside these discs we will safely use
the small $\e$ limits of the correlation functions given in the appendix while
inside the discs we'll explore (exact in $\e$) OPE. We will see that the trace
of the intermediate scale $l$ will be washed out entirely from the final
result. In present case the $\e=0$ limit of the four-point function is
given by (see appendix \ref{C})
\bea
\label{G0eps}
&&\langle \phi(x)\phi(0)\phi(1)\phi(\infty)\rangle\\
&&=\left|\frac{1-2x+3 x^2-2x^3+\frac{x^4}{3}}
{x^2(1-x)^2}\right|^2 +
\frac{16}{3}\left|\frac{1-\frac{3x}{2}+x^2-\frac{x^3}{4}}
{x(1-x)^2}\right|^2
+\frac{5}{9}\left|\frac{x}{1-x}\right|^4\nonumber
\eea
With required accuracy $I(x)\approx \pi/\e$. It is convenient to carry
out the integration in radial coordinates $x=r \exp (i\varphi)$,
$\bar{x}=r \exp (-i\varphi)$, $d^2x=r dr d\varphi$. The result
of integration over the angular variable $\varphi$ will depend
on the region where the radial coordinates takes its value
\bea
\int R(x,\bar{x})d\varphi=\left\{
\begin{array}{c}
Res_{x=0}R(x,r^2/x)/x+Res_{x=r^2}R(x,r^2/x)/x,\quad if\,\, r<1\\
Res_{x=0}R(x,r^2/x)/x+Res_{x=1}R(x,r^2/x)/x,\quad if\,\, r>1
\end{array}
\right.
\eea
for arbitrary rational function $R(x,\bar{x})$ with poles located
at $x=0$ or $x=1$.
In particular when $R(x,\bar{x})$ is the r.h.s. of the eq. (\ref{G0eps})
we get
\bea
\frac{3 r^{10}-9 r^8+25 r^6-23 r^4+7 r^2+3}{3 r^4 (1-r^2)^3}\, ,\quad &if&\,\, r<1\nonumber\\
\frac{3 r^{10}+7 r^8-23 r^6+25 r^4-9 r^2+3}{3 r^4 (r^2-1)^3}\, ,\quad &if&\,\, r>1
\eea
After performing the remaining elementary integration over $r$ we finally get
\bea
\int_{\Omega_{l,l_0}}\langle \phi(x)\phi(0)\phi(1)\phi(\infty)\rangle d^2x =
\frac{2 \pi }{l^2}+\frac{\pi }{2l_0^2}-\frac{33 \pi }{8}-\frac{32 \pi}{3} \log (2l_0l^2)+\cdots
\label{Omegaint}
\eea
where (see Fig. \ref{Fig1} )
\[
\Omega_{l,l0}=(D_{1-l_0,0}\backslash D_{l,0})\cup (D_{1/l,0}\backslash D_{1+l_0,0})
\]
and the dots stand for negligible terms of order $l$ or $l_0/l$.
There is a subtlety to be treated carefully here. The fact that the
part of the white narrow ring (of width $2l_0$) outside of the red
circle  is missing from the integration region $\Omega_{l,l_0}$ is
insignificant since its inclusion would produce only negligible terms
of order $l_0$. Instead we have to subtract the contribution of two
lens-like regions of $\Omega_{l,l_0}$ included in the red circle
(as already said, the contribution coming from the regions around
the singular points will be computed separately exploring OPE).\\
a) {\it{Integration over lens-like regions}}\\
Expanding (\ref{G0eps}) around $x\sim 1$ we get
\bea
\frac{1}{|x-1|^4}+\frac{2}{(x-1)^2}+\frac{2}{(\bar{x}-1)^2}
+\frac{16}{3|x-1|^2}+\cdots
\eea
where only the singular terms, whose integrals over the
region around $x=1$ diverge, are presented. The integrals of such
terms have been evaluated in Appendix \ref{D}. As  a result the contribution
of the lens-like regions, to be subtracted from the r.h.s. of eq.
(\ref{Omegaint}), is equal to
\bea
\frac{\pi}{\e}  \left(\left(-\frac{\pi }{l^2}-\frac{\pi }{8}\right)+
\frac{\pi }{2 l_0^2}
+2\times 2 \pi +\frac{16}{3} \left(2 \pi  \log \frac{l}{2 l_0}\right)
\right)
\label{lensint}
\eea
b) {\it{Contributions of the regions around singularities}}\\
It remains to calculate the contributions of the regions $D_{l,0}$,
$D_{l,1}$, $D_{l,\infty}$ to the integral (\ref{2ndorderintreduced}).
Evidently the first two regions give identical contributions, so let's concentrate on
the region $D_{l,0}$ for definiteness. To calculate the four-point function
in this region we apply the OPE (all the structure constants we use in this
paper can be extracted from the general formula (\ref{StrConstGen}))
\bea
\phi(x)\phi(0)=(x\bar{x})^{-2\D}(1+\cdots)+
C_{(1,3)(1,3)}^{(1,3)}(x\bar{x})^{-\D}\left(\phi(0)+\cdots\right)
\label{phiphiOPE}
\eea
Taking into account (\ref{I0e120lim}) and that
\bea
C_{(1,3)(1,3)}^{(1,3)}C_{(1,3)(1,3)(1,3)}=\frac{16}{3}\,(1-3 \epsilon+O(\e^2))
\eea
one easily gets
\bea
&&\int_{D_{l,0}\backslash D_{l_0,0}} I(x) \langle \phi(x)\phi(0)\phi(1)\phi(\infty)
\rangle d^2x\nonumber\\
&&\approx \frac{\pi^2}{\e l_0^{2-4\e}}-\frac{\pi^2}{\e l^2}+
\frac{32 \pi ^2 (\log (l)-3)}{3 \epsilon }+\frac{32 \pi ^2}{3 \epsilon ^2}
\label{0contrib}
\eea
where the first two terms come from the identity and the last two terms from the
$\phi$ field channel of OPE (\ref{phiphiOPE}) respectively.\\
c) {\it{Contribution of}} $x\sim \infty$\\
At large $x$ we first make use of eq. (\ref{Ginf}) to pass to the inverse
variable $1/x\sim0$ and then we apply OPE. The calculation
is similar to the previous case, the main difference being the fact
that at this limit $I(x)$ becomes simply $\frac{\pi}{\e} (x\bar{x})^{-2\e}$.
The result is
\bea
&&\int_{D_{l,\infty}\backslash D_{l_0,\infty}} I(x) \langle \phi(x)\phi(0)\phi(1)\phi(\infty)
\rangle d^2x\nonumber\\
&&\approx \frac{\pi^2}{\e l_0^{2-4\e}}-\frac{\pi^2}{\e l^2}+
\frac{16 \pi^2}{3}\left(\frac{1}{\e^2}-\frac{3-2\log l}{\e}
\right)
\label{infcontrib}
\eea
Let's pick up all the ingredients: (\ref{Omegaint}) times
$I(x)\approx \frac{2\pi}{\e}$, minus (\ref{lensint}),
plus twice (\ref{0contrib}), and plus (\ref{infcontrib}). We get
\bea
\frac{3 \pi ^2}{l_0^{2-4\e}}-\frac{88 \pi ^2}{\epsilon }
+\frac{80 \pi ^2}{3 \epsilon ^2}+O(\e^0)
\eea
As expected, the $l$ dependence disappeared. The presence of
the divergent term $\frac{3 \pi ^2}{l_0^{2-4\e}}$ also is not surprising.
In our naive regularization scheme we could cancel this infinity
by adding an appropriate, proportional to the area
counter-term in action. In fact, if we would have been able to treat
the integral (\ref{2ndorderintreduced}) analytically as continuation
from a region of parameters where integral converges, then this
kind of non-analytic divergent term couldn't emerge at all. In what follows
we'll simply drop out such terms without further ado.\\
Thus for the two point function
$G(x,\l)=\langle\phi(x)\phi(0)\rangle_\l$ we get (c.f. (\ref{2pfunclam}))
\bea
G(x,\lambda )&=&(x\bar{x})^{-2+2\e}\left(1-\l\,\frac{4\pi}{\sqrt{3}}\,
\left(\frac{2}{\e}-3+O(\e)\right)(x\bar{x})^\e\right.\nonumber\\
&+& \left.\frac{\l^2}{2}
\left(\frac{80\pi^2}{3\e^2}-\frac{88\pi^2}{\e}+O(\e^0)\right)
(x\bar{x})^{2\e}+\cdots\right)
\eea
Following A.~Zamolodchikov let us introduce a new coordinate $g$
("renormalized" coupling constant) in the space
of one-parameter family of theories (\ref{action}) instead of the initial
coupling $\l$ and introduce the local field $\phi^{(g)}=\partial_g
{\cal H}$ (according to (\ref{action}) the initial "bare" perturbing
field  $\phi=\partial_\l {\cal H}$). The new coupling $g$ is
fixed by the requirement that
the two point function $G(x,g)=\langle\phi^{(g)}(x)\phi^{(g)}(0)\rangle_\l$
satisfies the normalization condition
\bea
G(1,g)=1
\label{2pnorm}
\eea
Then the $\beta$-function can be computed from the identity
(see \cite{Zamolodchikov:1987})
\bea
\Theta(x)=\e\l\phi(x)=\beta(g)\phi^{(g)}(x)
\label{Theta}
\eea
where $\Theta$ is the trace of the energy-momentum tensor.
Combining (\ref{2pnorm}) and (\ref{Theta}) one easily finds
\bea
\partial_\l g=\sqrt{G(1,\l)}
\label{dg}
\eea
and
\bea
\beta(g)=\e\l\sqrt{G(1,\l)}
\label{betaexact}
\eea
The equation (\ref{dg}) allows one to express $g$ in terms of $\l$
(the integration constant can be set to zero so that the
unperturbed CFT will corresponds to $g=0$)
\bea
g=\l-\frac{\pi  \l^2}{\sqrt{3}} \left(\frac{2}{\e}-3+O(\e)\right)
+\frac{2 \pi ^2 \l^3}{3} \left(\frac{2}{\e^2}-\frac{7}{\e}+O(1)
\right)+O(\l^4)
\label{gl}
\eea
or, inversely
\bea
\l=g+\frac{\pi  g^2}{\sqrt{3}} \left(\frac{2}{\e}-3+O(\e)\right)
+\frac{2 \pi ^2 g^3}{3} \left(\frac{2}{\e^2}-\frac{5}{\e}+O(1)
\right)+O(g^4)
\label{lg}
\eea

Inverting and replacing
in (\ref{betaexact}) $\l$ in favour of $g$ we get
\bea
\beta(g)=\e g -\frac{\pi g^2}{\sqrt{3}}\,\left(2-3\e+O(\e^2)\right)
-\frac{4\pi^2g^3}{3}
\left(1+O(\e)\right)+\cdots
\eea
The equation
\bea
\beta(g^*)=0
\eea
admits a nonzero solution
\bea
2\pi g^*=\sqrt{3}\,\epsilon +\frac{\sqrt{3}}{2}\, \epsilon ^2
+O(\e^3)
\label{gfp}
\eea
so we have a non-trivial infrared fixed point. In \cite{Zamolodchikov:1987} this
fixed point has been identified with the minimal model $M_{p-1}$
and the local field $\phi^{(g^*)}$ with the field $\phi_{3,1}^{(p-1)}$.
Now we are in a position to check this identification more
accurately. The anomalous dimension of $\phi^{(g^*)}$ is related
to the slope of $\beta$-function
\bea
\D^*=1-\partial_g \beta(g)\left|_{g=g^*}=1+\e+\e^2+O(\e^3)\right.
\eea
which matches to the conformal dimension of $\phi_{3,1}^{(p-1)}$
computed from the Kac formula (\ref{DimGen}). Also the shift of the
central charge \cite{Zamolodchikov:1986}
\bea
 c^*-c_p=-12 \pi^2\int_0^{g^*}\beta(g)dg=
-\frac{3 \epsilon ^3}{2}-\frac{9 \epsilon ^4}{4}+O(\e^5)
\eea
neatly matches to the exact expression
\[
c_{p-1}-c_p=-\frac{12}{p(p^2-1)}=
-\frac{3 \epsilon ^3}{(2-\epsilon) (1-\epsilon )}.
\]


\section{Field renormalization and the UV - IR map}
\label{FRUIM}
In this section we calculate the matrices of anomalous dimensions for several
classes of fields. Diagonalization of these matrices at the IR fixed point
provides a detailed map between the UV local fields and their image under
RG flow in the IR theory.

\subsection{Primary fields $\phi_{n,n}$}
\label{PFnn}
This is the simplest case to analyze since the fields $\phi_{n,n}$ never
get mixed with other fields \cite{Zamolodchikov:1987}. This follows from
the structure of the OPE involving the perturbing field $\phi_{1,3}$.
The subspace of fields which is generated by the field $\phi_{n,n}$ and is
closed w.r.t. OPE with $\phi_{1,3}$, doesn't contain any other field
with a dimension close to $\D_{n,n} =O(\e^2)$. \\
We are going to calculate corrections to the anomalous dimension up
to the order $\e^4$. That is why for the present purpose the knowledge
of the four point function\\
$\langle\phi(x)\phi_{n,n}(0)\phi_{n,n}(1)\phi(\infty)\rangle$ up to
$\e^2$ correction is required. As in previous case, to find this
correlation function we first used AGT relation to find the relevant conformal
blocks up to sufficiently large level (actually the computations
were performed up to the order $x^6$ terms). Expanding a conformal block
up to $\e^2$ and examining first few coefficients of the resulting
power series in $x$ it is possible to guess the entire power series and
identify it with some elementary function. Having in our disposal the
expression for the correlation function we then checked that it satisfies
all the nontrivial physical requirements: the single-valuedness and
the compatibility with OPE around the points $x\sim 1$ and $x\sim \infty$.
Here is the final expression (see Appendix \ref{C})
\bea
&&\langle \phi(x)\phi_{n,n}(0)\phi_{n,n}(1)\phi(\infty)\rangle\\
&&=1+\frac{\e^2(n^2-1)}{12}\left(\frac{1}{2x(x-1)}+
\frac{1}{2\bar{x}(\bar{x}-1)}+4\log ^2\left|\frac{1-x}{x}\right|\right)
+O(\e^3)
\nonumber
\label{Gneps}
\eea
From eq. (\ref{I0eps12}), up to order $\e$, $I(x)$ is equal to
\bea
I(x)=\frac{\pi}{\epsilon} - 2\pi \log|(1 - x)x|+
8 \pi\e  \log|x|\log|1 - x|+O(\e^2)
\eea
Now we are ready to perform integration over the region $\Omega_{l_0,l}$
(see Fig. \ref{Fig1}). Since the singularities at $x\sim 0$ and $x\sim 1$
are integrable, we can put $l_0=0$. As in Section \ref{bf} the integration over
the angular variable should be performed separately for the cases $0<|x|<1$
and $|x|>1$. Integration of rational expressions we have already discussed
earlier. As about the logarithmic terms, they can be easily handled first
expanding into power series in $x$ if $|x|<1$ or in $1/x$
if $|x|>1$.
Then we proceed with the radial integration. Both steps are elementary
and we present only the final result:
\bea
\label{Omegaintphin}
&&\int_{\Omega_{l,l_0}}I(x)\langle \phi(x)\phi_{n,n}(0)
\phi_{n,n}(1)\phi(\infty)\rangle d^2x \approx
\frac{\e \pi^2 (n^2-1)}{3}\left(\log\frac{1}{l}
+1\right)\\
&&+\frac{\pi^2}{l^2}\,(2+4\e+(4+8\e)\log l+8\e\log^2 l)-\pi^2(1+4\e)
-\frac{\e \pi^2(n^2-1)}{12}+\frac{\pi^2}{\e l^2}\nonumber
\eea
Due to the already mentioned mildness of singularities at $0$ and $1$
the only remaining contribution to be taken into account comes from
the neighbourhood of $\infty$ \ie from
$D_{l,\infty}\backslash D_{l_0,\infty}$.\\
At large $x$ it is convenient to employ eq. (\ref{Ginf})
\bea
\langle \phi(x)\phi_{n,n}(0)
\phi_{n,n}(1)\phi(\infty)\rangle=(x\bar{x})^{-2\D}
\langle \phi(1/x)\phi(0)
\phi_{n,n}(1)\phi_{n,n}(\infty)\rangle
\label{Gninf}
\eea
and apply the OPE (\ref{phiphiOPE}) with $x$ replaced by $1/x$.
The correlation function decomposes into a sum of two partial
amplitudes one corresponding to the identity and the other to
the field $\phi$.\\
a) {\it{Contribution of identity}}\\
The prefactor $(x\bar{x})^{-2\D}$ in (\ref{Gninf}) compensates
the factor $(x\bar{x})^{2\D}$ accompanying the identity operator in OPE
and with sufficient accuracy we can replace this partial amplitude
by $1$. It is straightforward to expand $I(x)$ given by
(\ref{I0eps12}) at large $x$ keeping only those terms which after
integration may produce non-vanishing terms in small $l$ limit
\bea
I(x)\approx \frac{\pi \gamma^2 (2 \epsilon)}{\gamma (4 \epsilon )} (x \bar{x})^{-2 \e}
\left(1+\frac{\e}{x}\right)\left(1+\frac{\e}{\bar{x}}\right)+
\frac{\pi \gamma (4 \epsilon -1)}{\gamma (2 \epsilon )^2} (x \bar{x})^{2 \e-1}
\eea
Integrating this expression over the region $D_{l,\infty}\backslash D_{l_0,\infty}$,
dropping out, as earlier, all singular in $l_0$ terms and expanding
the result up to the linear in $\e$ terms we get
\bea
-\frac{\pi ^2}{\e l^2}+\pi ^2-\frac{2 \pi ^2 (2\log (l)+1)}{l^2}+
\frac{4\pi^2 \e \left( l^2-2 \log ^2(l)-2 \log (l)-1\right)}{l^2}
\label{Gninfident}
\eea
b) {\it{Contribution of the field}} $\phi_{1,3}$\\
This contribution is
\bea
\int_{D_{l,\infty}\backslash D_{l_0,\infty}}
\frac{\pi}{\e}\, C^2_{(1,3)(n,n)(n,n)}\,(x\bar{x})^{-2\e -2+2\e+1-\e}d^2x
\eea
where $\frac{\pi}{\e} (x\bar{x})^{-2\e}$ is just the function
$I(x)$ with required accuracy, $(x\bar{x})^{-2+2\e}$ is the prefactor of (\ref{Gninf}),
$(x\bar{x})^{1-\e}$ comes from OPE and the squared structure constant
is equal to
\bea
C^2_{(1,3)(n,n)(n,n)}=\frac{\e^2(1+\e)(n^2-1)}{6}+O(\e^4)
\eea
The integral is converging at the limit $l_0\rightarrow 0$,
so we may perform integration over the entire region $D_{l,\infty}$.
The result reads
\bea
\frac{\pi^2(n^2-1)(1-\e +2\e \log(l))}{6}
\label{Gninfphi}
\eea
The sum of all contributions (\ref{Omegaintphin}), (\ref{Gninfident}),
and (\ref{Gninfphi}) is
\bea
\frac{\pi^2\left(n^2-1\right)(2+5\epsilon)}{12}+O(\e^2)
\label{PhnPhnC2}
\eea
Combining this with the first order in coupling constant
contribution
\bea
\int \langle \phi_{n,n}(1)\phi_{n,n}(0)\phi(x)\rangle d^2x=
\frac{\pi  \left(n^2-1\right)(2+5 \epsilon )\,\epsilon }{8 \sqrt{3}}
+O(\e^4)
\eea
where the value
\bea
C_{(1,3)(n,n)(n,n)}=\frac{\left(n^2-1\right)
(2+5\epsilon)\,\e^2 }{16\sqrt{3}}+O(\e^4)
\eea
for the structure constant is inserted,
we get
\bea
G_n(x,\lambda )&\equiv &\langle \phi_{n,n}(x)\phi_{n,n}(0)\rangle_\l
\nonumber\\
&=&(x\bar{x})^{-2\D_{n,n}}\left(1-\l\,\frac{\pi  \left(n^2-1\right)
(2+5 \epsilon+O(\e^2) )\,\epsilon }{8 \sqrt{3}}\,
(x\bar{x})^\e \right.\nonumber\\
&+&\left.\frac{\l^2}{2}
\frac{\pi^2\left(n^2-1\right)(2+5\epsilon+O(\e^2))}{12}\,
(x\bar{x})^{2\e}+\cdots\right)
\eea
Let's introduce the renormalized field $\phi_{n,n}^{(g)}=B(\l)\phi_{n,n}$ by
requiring that the two point function $G_n(x,g)=\langle\phi_{n,n}^{(g)}(x)\phi_{n,n}^{(g)}(0)\rangle_\l$
satisfies the normalization condition
\bea
G_n(1,g)=1
\label{2pnnorm}
\eea
so that
\bea
B(\l)=\frac{1}{\sqrt{G_n(1,\l)}}
\eea
Then for the anomalous dimension we get (cf.
eq. (\ref{Dimlamb}), derived for a more general situation)
\bea
\D^{(g)}_{n,n}=\D_{n,n}+\e \l\, \partial_\l \log B=
\D_{n,n}-\frac{\e \l}{2}\, \partial_\l G_n(1,\l)
\eea
In view of (\ref{lg}) we find
\bea
\D^{(g)}_{n,n}&=&\D_{n,n}\\
&+&\frac{\pi  g \left(n^2-1\right)
\epsilon ^2 \left(2+5 \epsilon+O(\e^2)\right)}{16 \sqrt{3}}
-\frac{\pi^2 g^2 \left(n^2-1\right)
\epsilon ^2(1+O(\e))}{8}+O(g^3)\nonumber
\eea
So that at the fixed point
\bea
\D^{(g^*)}_{n,n}=\frac{(n^2-1)(4\e^2+6\e^3+7\e^4+O(\e^5))}{64}
\eea
which completely agrees with the dimension $\D_{n,n}^{(p-1)}$
of the field $\phi_{n,n}^{(p-1)}$ in the minimal model $M_{p-1}$.
Thus the conclusion of A.~Zamolodchikov that under RG the UV
field $\phi_{n,n}^{(p)}$ flows to IR $\phi_{n,n}^{(p-1)}$ is robust
also against our second order test.


\subsection{Renormalization of the fields $\phi_{n,n+1}$ and $\phi_{n,n-1}$}
\label{Pnp1nm1}
Already in this case one encounters with the phenomenon of mixing.
The OPE  $\phi_{1,3}\phi_{n,n+1}$ produces besides $\phi_{n,n+1}$ also the primary field $\phi_{n,n-1}$, both having dimensions close to $1/4$ in large $p$ limit.
Thus we have to consider  the correlation functions\\ $\langle\phi(x)\phi_{n,n\pm 1}(0)\phi_{n,n\pm 1}(1)\phi(\infty)\rangle$ with all four possible choices of signs.
The strategy is exactly the same as in previous sections and for each choice we
will follow the steps performed in Section \ref{PFnn}.\\
\subsubsection{ Correlation function $\langle\phi_{n,n+1}(1)\phi_{n,n+ 1}(0)\rangle_\l$}
a) {\it{Contribution of the region} $\Omega_{l,l_0}$}\\
This is given by the integral
\bea
\int_{\Omega_{l,l_0}} I(x)\langle\phi(x)\phi_{n,n+1}(0)\phi_{n,n+ 1}(1)\phi(\infty)\rangle d^2x
\label{Omegaintnp1np1}
\eea
The large $p$ limit of the four-point function found from AGT relation is
(See Appendix \ref{C}):
\bea
\langle \phi(x)\phi_{n,n+1}(0)\phi_{n,n+1}(1)\phi(\infty)\rangle
=\frac{2 (n+2)}{3n} \left|\frac{x-\frac{1}{2}}{x(x-1)}\right|^2+\left|\frac{x^2-x+\frac{1}{2}}
{x(x-1)}\right|^2+O(\e)\nonumber\\
\label{eps0np1np1}
\eea
With required accuracy $I(x)$ can be replaced by $\frac{\pi}{\epsilon}$.
The integral (\ref{Omegaintnp1np1}) can be performed using the technique
already explored in computing (\ref{Omegaint}) or (\ref{Omegaintphin}).
The result is
\bea
&&\int_{\Omega_{l,l_0}} I(x)\langle\phi(x)\phi_{n,n+1}(0)\phi_{n,n+ 1}(1)\phi(\infty)\rangle d^2x\nonumber\\
&&\approx \frac{\pi }{l^2}-\pi-\frac{\pi  (13 n+20) \log (l)}{6 n}-\frac{\pi  (5 n+4) \log \left(2 l_0\right)}{6 n}
\label{np1np1Omegacont}
\eea
b) {\it{Contribution of lens-like regions}} \\
Near $x\sim 1$ the r.h.c. of eq. (\ref{eps0np1np1}) becomes
$\frac{5n+4}{12n |x-1|^{2}}\,$, hence the contribution
of the lens-like regions near $x\sim 1$  to be subtracted from the r.h.s. of eq.
(\ref{np1np1Omegacont}) is equal to (see (\ref{LensLog}))
\bea
\frac{\pi^2 (5n+4)}{6n\e} \log \left(\frac{l}{2 l_0}\right)
\label{lensnp1np1}
\eea
c) {\it{Contributions of the regions}} $D_{l,0}$, $D_{l,1}$ {\it{and}} $D_{l,\infty}$\\
Contributions of $D_{l,0}$ and $D_{l,1}$ obviously are identical
so we will concentrate on $D_{l,0}$ only.
The relevant OPE is
\bea
\phi(x)\phi_{n,n+1}(0)&=&(x\bar{x})^{-\D}C_{(1,3)(n,n+1)}^{(n,n+1)}
(\phi_{n,n+1}+\cdots)\nonumber\\&+&
C_{(1,3)(n,n+1)}^{(n,n-1)}(x\bar{x})^{\D_{n,n-1}-\D_{n,n+1}-\D}
\left(\phi_{n,n-1}(0)+\cdots\right)
\label{phiphinp1OPE}
\eea
Taking into account that in this region $I(x)\approx \frac{\pi }{\epsilon }-\pi  \log \left(|x|^2\right)$ (see (\ref{I0e120lim})) and that
\bea
C_{(1,3)(n,n+1)(n,n+1)}^2&=&\frac{(n+2)^2 \left(1-(2 n-1) \epsilon\right)}{12 n^2}
+O(\e^2)\nonumber\\
C_{(1,3)(n,n+1)(n,n-1)}^2&=&\frac{\left(n^2-1\right) (1+\epsilon )}{3 n^2}
+O(\e^2)
\eea
we get
\bea
\int_{D_{l,0}\backslash D_{l_0,0}} I(x) \langle \phi(x)\phi(0)\phi(1)\phi(\infty)
\rangle d^2x
\approx \frac{\pi ^2 (n+2)^2}{6 n^2} \left(\frac{1}{\epsilon ^2}+\frac{1-2 n+\log l}{\epsilon }\right)\nonumber\\
+\frac{2 \pi ^2 \left(n^2-1\right)}{3 n^2 (n+2)^2} \left(\frac{n+4}{\epsilon ^2}+\frac{n+4+(n+2)^2 \log l}{\epsilon }\right)\qquad
\label{np1np10cont}
\eea
Above two terms come from two primaries $\phi_{n\pm1}$ appearing on the r.h.s.
of the OPE (\ref{phiphinp1OPE}).

The contribution from the region $D_{l,\infty}$ is completely analogous
to the case of the correlation function $\langle\phi\phi\rangle_\l$ discussed
in Section \ref{bf}. The only difference is that now the contribution of the field
$\phi$ which appears in u-channel OPE is proportional to
\bea
C_{(1,3)(1,3)}^{(1,3)}C_{(1,3)(n,n+1)(n,n+1)}=\frac{2 (n+2) \left(1-(n+1)
\epsilon \right)}{3 n}+O(\e^2)
\eea
instead of $C_{(1,3)(1,3)(1,3)}^2\approx \frac{16 (1-3\e)}{3}$. The result
is (c.f. eq. (\ref{infcontrib}))
\bea
&&\int_{D_{l,\infty}\backslash D_{l_0,\infty}} I(x) \langle \phi(x)\phi_{n,n+1}(0)\phi_{n,n+1}(1)\phi(\infty)
\rangle d^2x\nonumber\\
&&\approx -\frac{\pi^2}{\e l^2}+
\frac{2 \pi^2 (n+2)}{3n}\left(\frac{1}{\e^2}-\frac{n+1-2\log l}{\e}
\right)
\label{infcontnp1np1}
\eea
Picking up all the contributions: (\ref{np1np1Omegacont}) minus
(\ref{lensnp1np1}) plus twice (\ref{np1np10cont}) and plus (\ref{infcontnp1np1}), we get
\bea
\frac{\pi ^2 \left(3 n^3+24 n^2+64 n+44\right)}{3 n (n+2)^2 \epsilon ^2}
-\frac{4 \pi ^2 (n+1) \left(n^3+7 n^2+14 n+5\right)}{3 n (n+2)^2 \epsilon }
+O(\e^0)
\label{npm1C211}
\eea

\subsubsection{ Correlation function $\langle\phi_{n,n-1}(1)\phi_{n,n+1}(0)\rangle_\l$}
\label{CFnm1np1}
a) {\it{Contribution of the region}} $\Omega_{l,l_0}$\\
is given by the integral
\bea
\int_{\Omega_{l,l_0}} I(x)\langle\phi(x)\phi_{n,n+1}(0)\phi_{n,n- 1}
(1)\phi(\infty)\rangle d^2x
\label{Omegaintnm1np1}
\eea
The large $p$ limit of the four-point function now is (see Appendix \ref{C}):
\bea
\langle \phi(x)\phi_{n,n+1}(0)\phi_{n,n-1}(1)\phi(\infty)\rangle
=\frac{\sqrt{n^2-1}}{3n}\, \left|\frac{2x-1}{x(x-1)}\right|^2+O(\e)
\label{eps0nm1np1}
\eea
Using the last equality in (\ref{Ieps}) we see that $I(x)\approx \frac{16 \pi }{\left(16-n^2\right) \epsilon }$ and
for the result of the integral (\ref{Omegaintnm1np1}) we get
\bea
&&\int_{\Omega_{l,l_0}} I(x)\langle\phi(x)\phi_{n,n+1}(0)\phi_{n,n-1}(1)\phi(\infty)\rangle d^2x\nonumber\\
&&\approx \frac{32 \pi ^2 \sqrt{n^2-1} \left(5 \log (l)+
\log \left(2l_0\right)\right)}{3 n \left(n^2-16\right) \epsilon }
\label{nm1np1Omegacont}
\eea
b) {\it{Contribution of lens-like regions}} \\
Near $x\sim 1$ the r.h.s. of eq. (\ref{eps0nm1np1}) behaves as
$\frac{\sqrt{n^2-1}}{3n}\, |x-1|^{-2}$ and the contribution
of the lens-like regions near $x\sim 1$, which should be subtracted
from the r.h.s. of eq. (\ref{nm1np1Omegacont}) is (see (\ref{LensLog}))
\bea
\frac{32 \pi^2\sqrt{n^2-1}}{3n(16-n^2)\e} \log \left(\frac{l}{2 l_0}\right)
\label{lensnm1np1}
\eea
c) {\it{Contributions of the regions}} $D_{l,0}$, $D_{l,1}$ {\it{and}} $D_{l,\infty}$\\
We will treat contributions of $D_{l,0}$ and $D_{l,1}$ separately.\\
i) $D_{l,0}$ {\it{contribution}}. \\
The relevant OPE is
\bea
\phi(x)\phi_{n,n+1}(0)&=&C_{(1,3)(n,n+1)}^{(n,n+1)}(x\bar{x})^{-\D}
(\phi_{n,n+1}+\cdots)\nonumber\\&+&
C_{(1,3)(n,n+1)}^{(n,n-1)}(x\bar{x})^{\D_{n,n-1}-\D_{n,n+1}-\D}
\left(\phi_{n,n-1}(0)+\cdots\right)\nonumber
\eea
It follows from (\ref{Ieps}) that in this region with sufficient accuracy
\[
I(x)\approx \frac{8 \pi}{n\e}  \left(-\frac{1}{n+4}
-\frac{(x \bar{x})^{\D_{n,n+1}-\D_{n,n-1}}}{n-4}\right)
\]
From (\ref{StrConstGen})
\bea
C_{(1,3)(n,n+1)}^{(n,n+1)}C_{(1,3)(n,n+1)(n,n-1)}=\frac{\sqrt{n^2-1} \,(n+2)(1-(n-1)\epsilon )}{6 n^2}
+O(\e^2)\nonumber\\
C_{(1,3)(n,n+1)}^{(n,n-1)}C_{(1,3)(n,n-1)(n,n-1)}=\frac{\sqrt{n^2-1} \,(n-2)(1+(n+1)\epsilon )}{6 n^2}
+O(\e^2)
\label{nm1np1SC0}
\eea
and for the $D_{l,0}$ contribution we get
\bea
&&\int_{D_{l,0}\backslash D_{l_0,0}} I(x) \langle \phi(x)\phi(0)\phi(1)\phi(\infty)
\rangle d^2x\nonumber\\
&&\approx -\frac{4 \pi ^2 (n+2) \sqrt{n^2-1} ((n-8) (1-(n-1) \epsilon )+4 (n-2)
\epsilon  \log l)}{3 (n^2-16) (n-2) n^2 \epsilon ^2}\nonumber\\
&&-\frac{4 \pi ^2 (n-2) \sqrt{n^2-1} ((n+8) (1+(n+1) \epsilon )+4 (n+2) \epsilon
\log l)}{3 (n^2-16) n^2 (n+2) \epsilon ^2}
\label{nm1np10cont}
\eea
where two terms correspond to the two intermediate primaries $\phi_{n\pm1}$.\\
ii) $D_{l,1}$ {\it{contribution}}.\\
The relevant OPE is
\bea
\phi(x)\phi_{n,n-1}(1)=
C_{(1,3)(n,n-1)}^{(n,n+1)}((x-1)(\bar{x}-1))^{\D_{n,n+1}-\D_{n,n-1}-\D}
(\phi_{n,n+1}+\cdots)\nonumber\\+
C_{(1,3)(n,n-1)}^{(n,n-1)}((x-1)(\bar{x}-1))^{-\D}
\left(\phi_{n,n-1}(1)+\cdots\right)\nonumber
\eea
Considering $x\rightarrow 1$ limit of (\ref{Ieps}) we get
\[
I(x)\approx \frac{8 \pi}{n\e}  \left(-\frac{1}{n-4}
-\frac{((x-1) (\bar{x}-1))^{\D_{n,n-1}-\D_{n,n+1}}}{n+4}\right)
\]
The combinations of structure constants relevant for this case
are those already presented in (\ref{nm1np1SC0}). For the $D_{l,1}$ contribution we get
\bea
&&\int_{D_{l,1}\backslash D_{l_0,1}} I(x) \langle \phi(x)\phi(0)\phi(1)\phi(\infty)
\rangle d^2x\nonumber\\
&&\approx -\frac{4 \pi ^2 (n+2) \sqrt{n^2-1} ((n-8) (1-(n-1) \epsilon )
+4 (n-2) \epsilon  \log l)}{3 (n^2-16) (n-2) n^2 \epsilon ^2}\nonumber\\
&&-\frac{4 \pi ^2 (n-2) \sqrt{n^2-1} ((n+8) (1+(n+1) \epsilon)
+4 (n+2) \epsilon  \log (l))}{3 (n^2-16) n^2 (n+2) \epsilon ^2}
\label{nm1np11cont}
\eea
Again the two terms correspond to two intermediate primaries $\phi_{n\pm1}$.
Notice that due to some subtle interplay among quantities involved, for
the contribution of $D_{l,1}$ we got exactly the same result as for the contribution of $D_{l,0}$.\\
iii) $D_{l,\infty}$ {\it{contribution}}. \\
Since the structure constant $C_{(1,1)(n,n-1)(n,n+1)}=0$
only the field $\phi$ which appears in the u-channel OPE gives
a nonzero contribution. This contribution is proportional to
\bea
C_{(1,3)(1,3)}^{(1,3)}C_{(1,3)(n,n-1)(n,n+1)}=
\frac{4\sqrt{n^2-1}\,(1-\epsilon )}{3n}
+O(\e^2)
\eea
Approximating $I(x)$ by
$I(x)\approx \frac{\pi }{\epsilon }\,|x|^{-4 \epsilon }$
we get
\bea
\int_{D_{l,\infty}\backslash D_{l_0,\infty}} I(x) \langle \phi(x)\phi_{n,n+1}(0)\phi_{n,n+1}(1)\phi(\infty)
\rangle d^2x\nonumber\\
\approx -\frac{64 \pi ^2 \sqrt{n^2-1} (1-\epsilon  (1-2 \log l))}
{3 n \left(n^2-16\right) \epsilon ^2}
\label{infcontnm1np1}
\eea
It remains to collect all the contributions together to get
\bea
-\frac{16 \pi ^2 \sqrt{n^2-1}\, \left(5 n^2-44+
\left(n^2+20\right) \epsilon\right)}{3n \left(n^2-16\right)
\left(n^2-4\right) \epsilon ^2}+O(\e^0)
\label{npm1C221}
\eea
\subsubsection{The matrix of anomalous dimensions}
\label{MAD}
There is no need to calculate the remaining two point functions $\langle\phi_{n,n+1}(1)\phi_{n,n- 1}(0)\rangle_\l$ and $\langle\phi_{n,n-1}(1)\phi_{n,n-1}(0)\rangle_\l$ since the former is identical with $\langle\phi_{n,n-1}(1)\phi_{n,n+1}(0)\rangle_\l$ and the latter can be obtained from $\langle\phi_{n,n+1}(1)\phi_{n,n+1}(0)\rangle_\l$ by simply replacing $n\rightarrow -n$.
For simplicity of notation let us denote $\phi_{n,n+1}\equiv \phi_1$ and
$\phi_{n,n-1}\equiv \phi_2$, then the two-point functions can be represented as
\bea
\label{npm12pf}
G_{\a,\b}(x,\lambda )&\equiv& \langle \phi_{\a}(x)\phi_{\b}(0)\rangle_{\l}\\
&=&(x\bar{x})^{-\D_a-\D_{\b}}\left(\d_{\a,\b}-\l C^{(1)}_{\a,\b}
(x\bar{x})^{\e}+\frac{\l^2}{2}C^{(2)}_{\a,\b}(x\bar{x})^{2\e}+\cdots\right)
\nonumber
\eea
The first order coefficients $C^{(1)}_{\a,\b}$ are given by
\bea
C^{(1)}_{\a,\b}&=&\int \langle \phi_{\a}(1)\phi_{\b}(0)\phi(x)\rangle d^2x
\nonumber\\
&=& C_{(1,3)(\a)(\b)}\,\frac{\pi  \gamma (\epsilon +\D_{\a}-\D_{\b}) \gamma (\epsilon +\D_{\b}-\D_{\a})}
{\gamma (2 \epsilon )}
\label{C1}
\eea
From eq. (\ref{DimGen}) for the dimensions we have
\bea
\D_1&\equiv &\D_{n,n+1}=\frac{1}{4}-\left(\frac{n}{4}
+\frac{1}{8}\right) \epsilon +\frac{1}{16} \left(n^2-1\right) \epsilon ^2
+O(\e^3)\nonumber\\
\D_2&\equiv &\D_{n,n-1}=\frac{1}{4}+\left(\frac{n}{4}-\frac{1}{8}\right) \epsilon
+\frac{1}{16} \left(n^2-1\right) \epsilon ^2+O(\e^3)
\eea
Explicitly, up to $O(\e)$ terms we get
\bea
C^{(1)}_{1,1}&=&\frac{\pi  (n+2) (2-(2 n-1) \epsilon)}{2 \sqrt{3}\, n \epsilon }
+O(\e);
\,\, C^{(1)}_{2,2}=\frac{\pi  (n-2) (2+(2 n+1) \epsilon)}
{2 \sqrt{3}\, n \epsilon }+O(\e)\nonumber\\
C^{(1)}_{1,2}&=&C^{(1)}_{2,1}=-\frac{4 \pi  \sqrt{n^2-1} (\epsilon +2)}{\sqrt{3}\,n (n^2-4) \epsilon }+O(\e)
\label{npm1C1}
\eea
and for the second order coefficients we have (see (\ref{npm1C211}), (\ref{npm1C221}) )
\bea
C^{(2)}_{1,1}&=&\frac{\pi ^2 \left(3 n^3+24 n^2+64 n+44\right)}{3 n (n+2)^2 \epsilon ^2}
-\frac{4 \pi ^2 (n+1) \left(n^3+7 n^2+14 n+5\right)}{3 n (n+2)^2 \epsilon }
+O(\e^0)\nonumber\\
C^{(2)}_{2,2}&=&\frac{\pi ^2 \left(3 n^3-24 n^2+64 n-44\right)}{3 n (n-2)^2 \epsilon ^2}
+\frac{4 \pi ^2 (n-1) \left(n^3-7 n^2+14 n-5\right)}{3 n (n-2)^2 \epsilon }
+O(\e^0)\nonumber\\
C^{(2)}_{1,2}&=&C^{(2)}_{2,1}=-\frac{16 \pi ^2 \sqrt{n^2-1}\, \left(5 n^2-44+
\left(n^2+20\right) \epsilon\right)}{3n \left(n^2-16\right)
\left(n^2-4\right) \epsilon ^2}+O(\e^0)
\label{npm1C2}
\eea
Obviously the correlation function (\ref{npm12pf}) satisfies the
Callan-Symanzik equation
\bea
\left(x\partial_x+\D_{\a}+\D_{\b}-\e \l \partial_{\l} \right)
G_{\a,\b}(x,\lambda )=0
\label{CSlamb}
\eea
As in Section \ref{PFnn} let us introduce renormalized
fields
\[
\phi_{\a}^{(g)}=B_{\a,\b}(\l)\phi_{\b}
\]
and require that the two point functions $G_{\a,\b}^{(g)}(x)=\langle\phi_{\a}^{(g)}(x)\phi_{\b}^{(g)}(0)\rangle_\l$
satisfy the normalization condition
\bea
G_{\a,\b}^{(g)}(1)=\d_{\a,\b}
\label{npm1norm}
\eea
In matrix notations we may write
\bea
G^{(g)}(x)=B\cdot G(x)\cdot B^T
\eea
Comparing with (\ref{CSlamb}) we see that the renormalized two-point function
satisfies the equation
\bea
(x\partial_x-\b(g) \partial_{g})G_{\a,\b}^{(g)}+\sum_{\rho=1}^2(\G_{\a,\rho}G_{\rho,\b}^{(g)}
+\G_{\b,\rho}G_{\a,\rho}^{(g)})=0
\label{CSg}
\eea
where the $\beta$ function and the renormalized coupling $g$ have been
introduced in\\
Section \ref{bf} and the matrix of anomalous dimensions $\G$ is
defined as
\bea
\G=B{\hat\D}B^{-1}-\e \l B\partial_{\l}B^{-1}
\label{Dimlamb}
\eea
where
\bea
{\hat\D}=\left(
\begin{array}{cc}
\D_1&0\\0&\D_2
\end{array}
\right)
\eea
Expanding the matrix $B$ up to second order in $\l$
\bea
&&B=1+\l B_1+\l^2B_{2}+O(\l^3)\nonumber\\
&&B^{-1}=1-\l B_1+\l^2
(B_1^2-B_{2})+O(\l^3)
\eea
imposing the normalization condition (\ref{npm1norm}) and requiring
that the matrix of the anomalous dimensions (\ref{Dimlamb}) be symmetric,
we find
\bea
B_1&=&\frac{1}{2}\,C^{(1)}+\frac{1}{2\e}\,\left[{\hat \D},C^{(1)}\right]\nonumber\\
B_2&=&\frac{3}{8} \,\left(C^{(1)}\right)^2-\frac{1}{4}\,C^{(2)}
-\frac{1}{8\e}\,\left[{\hat \D},C^{(2)}\right]\nonumber\\
&+&\frac{1}{8\e}\,\left[\left[{\hat \D},C^{(1)}\right],C^{(1)}\right]
+\frac{1}{4\e}\,\left[{\hat \D},\left(C^{(1)}\right)^2\right]
+\frac{1}{8\e^2}\left[{\hat \D},C^{(1)}\right]^2
\label{npm1B}
\eea
Now all the ingredients to calculate the matrix of anomalous dimensions
(\ref{Dimlamb}) are at our disposal. Taking also into account the $\l$-$g$
relation (\ref{lg}), we get
\bea
\G_{1,1}&=&\D_1+\frac{\pi  g (n+2) (2-(2 n-1) \epsilon )}{4 \sqrt{3}\,n}
+\frac{\pi ^2 g^2}{2}\nonumber\\
\G_{1,2}&=&\G_{2,1}=\frac{\pi  g \sqrt{n^2-1}\,(2+\epsilon )}{2 \sqrt{3}\,n}
\nonumber\\
\G_{2,2}&=&\D_2+\frac{\pi  g (n-2) (2+(2 n+1) \epsilon )}{4 \sqrt{3}\,n}
+\frac{\pi ^2 g^2}{2}
\eea
Notice that all the matrix elements are regular at $\e=0$, all double and
single poles in $\e$  disappeared.
At the fixed point $g=g^*$ (see (\ref{gfp}))
\bea
\G_{1,1}^{(g^*)}&=&\frac{1}{4}-\frac{\left(2 n^2-n-4\right) \epsilon }{8 n}+\frac{\left(n^3-4 n^2+n+8\right) \epsilon ^2}{16 n}\nonumber\\
\G_{1,2}^{(g^*)}&=&\G_{2,1}^{(g^*)}=\frac{ \sqrt{n^2-1}\, \epsilon  (1+\epsilon)}{2n}\nonumber\\
\G_{2,2}^{(g^*)}&=&\frac{1}{4}+\frac{\left(2 n^2+n-4\right) \epsilon }{8 n}+\frac{\left(n^3+4 n^2+n-8\right) \epsilon ^2}{16 n}
\eea
It is easy to get the eigenvalues of this matrix
\bea
\D_1^{(g^*)}&=&\frac{1}{4}+\left(\frac{n}{4}
+\frac{1}{8}\right) \epsilon +\frac{1}{16} \left(n^2+4n+1\right) \epsilon ^2
\nonumber\\
\D_2^{(g^*)}&=&\frac{1}{4}-\left(\frac{n}{4}-\frac{1}{8}\right) \epsilon
+\frac{1}{16} \left(n^2-4n+1\right) \epsilon ^2
\eea
Up to $O(\e^3)$ terms they coincide with the dimensions $\D_{n+1,n}^{(p-1)}$
and $\D_{n-1,n}^{(p-1)}$ of the IR CFT $M_{p-1}$. We can easily identify
also the corresponding normalized eigenvectors and establish
the explicit map
\bea
\phi_{n+1,n}^{(p-1)}&=&\frac{1}{n}\,\phi_{1}^{(g^*)}+
\frac{\sqrt{n^2-1}}{n}\,\phi_{2}^{(g^*)}\nonumber\\
\phi_{n-1,n}^{(p-1)}&=&-\frac{\sqrt{n^2-1}}{n}\,\phi_{1}^{(g^*)}
+\frac{1}{n}\,\phi_{2}^{(g^*)}
\label{npm1map}
\eea
Remarkably the coefficients in (\ref{npm1map}) did not receive
neither $\e$ nor $\e^2$ corrections. Thus it is quite perceivable that
under the renormalization scheme (\ref{npm1norm}), which we have adopted
following A.~Zamolodchikov, the relation (\ref{npm1map}) is exact.
The same phenomenon we will encounter in the next section where
a more involved case of mixing of the three fields
$\phi_{n,n\pm 2}$ and $\partial{\bar \partial}\phi_{n,n}$ will be
considered.

\subsection{Renormalization of the fields $\phi_{n,n+2}$,
$\partial \bar{\partial}\phi_{n,n}$ and $\phi_{n,n-2}$}
\label{Pnp2nm2}
The OPE  $\phi_{1,3}\phi_{n,n+2}$ includes fields from the conformal families
$[\phi_{n,n+4}]$, $[\phi_{n,n+2}]$ and $[\phi_{n,n}]$. Similarly the
product $\phi_{1,3}\phi_{n,n-2}$ produces
fields from the families $[\phi_{n,n-4}]$, $[\phi_{n,n-2}]$ and
$[\phi_{n,n}]$. Since the dimensions
of the primary fields $\phi_{n,n\pm 2}$ and the descendant field $\partial \bar{\partial}\phi_{n,n}$ are close to $1$ in large $p$ limit, we
have a situation when these three fields effectively get
mixed along the RG flow\footnote{The fields $\phi_{n,n\pm 4}$ have larger dimensions $\sim 4$ and do not get mixed with these three fields.} \cite{Zamolodchikov:1987}. To find the matrix of anomalous  dimensions one has to calculate all the two point correlators of these fields.

\subsubsection{ Correlation function $\langle\phi_{n,n+2}(1)\phi_{n,n+ 2}(0)\rangle_\l$}
a) {\it{Contribution of the region}} $\Omega_{l,l_0}$\\
is given by the integral
\bea
\int_{\Omega_{l,l_0}} I(x)\langle\phi(x)\phi_{n,n+2}(0)\phi_{n,n+2}(1)\phi(\infty)\rangle d^2x
\label{Omegaintnp2np2}
\eea
At large $p$ from the AGT relation we have found that (see Appendix \ref{C})
\bea
\label{eps0np2np2}
&&\langle \phi(x)\phi_{n,n+2}(0)\phi_{n,n+2}(1)\phi(\infty)\rangle
=\left| \frac{3 x^4-6 x^3+9 x^2-6 x+1}{3 (x-1)^2 x^2} \right|^2\\
&&+\frac{8(3+n)}{3(1+n)}\left| \frac{(2 x-1)
\left(2 x^2-2 x+1\right)}{4 (x-1)^2 x^2} \right|^2+\frac{(3+n)(4+n)}{18n(n+1)}
\left|(x-1)^2 x^2 \right|^{-2}+O(\e)\nonumber
\eea
For present purposes $I(x)$ can be simply replaced by $\frac{\pi}{\epsilon}$. Performing the integration
(\ref{Omegaintnp2np2}) we get
\bea
\label{np2np2Omegacont}
&&\int_{\Omega_{l,l_0}} I(x)\langle\phi(x)
\phi_{n,n+2}(0)\phi_{n,n+2}(1)\phi(\infty)\rangle d^2x\\
&&\approx -\frac{16 \pi ^2 \left(2 n^2+5 n+1\right) \log (l)}{3 n (n+1) \epsilon }
-\frac{16 \pi ^2 (n+1) \log (l_0)}{3 n \epsilon }\nonumber\\
&&-\frac{\pi ^2 (66+33 n+128 (n+1) \log (2))}{24 n \epsilon }
+\frac{2 \pi ^2 (2 n+1)}{3 n l^2 \epsilon }
+\frac{\pi ^2 (n+2)}{6 n l_0^2 \epsilon }\nonumber
\eea
b) {\it{Contribution of lens-like regions}} \\
Expanding (\ref{eps0np2np2}) near $x\sim 1$ we get
\bea
&&\frac{2+n}{3n |x-1|^4}+\frac{2(n^2+n-2)(x+\bar{x}-2)}{3n(n+1) |x-1|^4}
\nonumber\\
&&+\frac{(n^2+5 n+6)\left((x-1)^2+(\bar{x}-1)^2\right)}{3 n (n+1) |x-1|^4}
+\frac{8(n+1}{3n |x-1|^2}
\eea
So the contribution of the lens-like regions near $x\sim 1$ is (see Appendix
\ref{D})
\bea
\frac{\pi}{\e}\left(\frac{2+n}{3n}\left(\frac{\pi}{2 l_0^2}-\frac{\pi}{l^2}-
\frac{\pi}{8}\right)
+\frac{2(n^2+n-2)\pi}{3n(n+1)}\right.\nonumber\\
\left.+\frac{ 2 \pi(n^2+5 n+6)}{3 n (n+1)}
+\frac{8(n+1)}{3n}\, 2 \pi\log \frac{l}{2l_0}\right)
\label{lensnp2np2}
\eea
c) {\it{Contributions of the regions}} $D_{l,0}$, $D_{l,1}$
{\it{and}} $D_{l,\infty}$\\
The contributions of $D_{l,0}$ and $D_{l,1}$ are identical
so we compute only the $D_{l,0}$ part.
Here are the relevant terms of the OPE
\bea
\phi(x)\phi_{n,n+2}(0)=(x\bar{x})^{-\D}C_{(1,3)(n,n+2)}^{(n,n+2)}
\phi_{n,n+2}+
C_{(1,3)(n,n+2)}^{(n,n)}(x\bar{x})^{\D_{n,n}-\D_{n,n+2}-\D}\nonumber\\
\times
\left(1+\frac{\D+\D_{n,n}-\D_{n,n+2}}{2 \D_{n,n}}\,xL_{-1} \right)\left(1+\frac{\D+\D_{n,n}-\D_{n,n+2}}{2 \D_{n,n}}\,\bar{x}\bar{L}_{-1} \right)\phi_{n,n}(0)\nonumber\\
+\cdots\quad
\label{phiphinp2OPE}
\eea
where $L_k$, $\bar{L}_k$ are the left and right Virasoro generators. To proceed let us notice that the effect of the Virasoro generator $L_{-1}$ or $\bar{L}_{-1}$ in the three point function is rather simple. Namely the relation
\bea
\langle L_{-1}\phi_{n,n}(0)\phi_{n,n+2}(1)\phi(\infty)\rangle=
\left(\D_{n,n}+\D_{n,n+2}-\D\right)\langle \phi_{n,n}(0)
\phi_{n,n+2}(1)\phi(\infty)\rangle\,\,\,\,
\label{Lm1PhnPhnp2Ph}
\eea
and a similar relation with $L_{-1}$ replaced by $\bar{L}_{-1}$ hold.
We see from (\ref{I0e120lim}) that in this region $I(x)$ can be approximated by
\bea
I(x)\approx \frac{\pi }{\epsilon }+\pi  \left(x+\bar{x}-\log |x|^2\right)
-2 \pi  \epsilon  (x+\bar{x}) \log |x|^2
\eea
We need also the combination of structure constants
\bea
&&C_{(1,3)(n,n+2)(n,n+2)}^2=\frac{4 (n+3)^2 (1-(n+2) \epsilon )}{3 (n+1)^2}
+O(\e^2)\nonumber\\
&&C_{(1,3)(n,n+2)}^{(n,n)}C_{(n,n)(n,n+2)(1,3)}=\frac{n+2}{3 n}
+O(\e^2)
\eea
Taking into account that
\bea
&&\D_{n,n}-\D_{n,n+2}-\D=-2+\frac{n+3}{2}\, \epsilon \nonumber\\
&&\frac{\left(\D+\D_{n,n}-\D_{n,n+2}\right)\left(\D_{n,n}
+\D_{n,n+2}-\D\right)}{2 \D_{n,n}}=-\frac{2 (n-1)}{n+1}+\frac{(n-1) (n+3) \epsilon }{2 (n+1)}
\nonumber
\eea
we get that the piece of integrand corresponding to the contribution of the family $[\phi_{n,n}]$ is equal to
\bea
\frac{2+n}{3n}\left|x\right|^{-4+(3+n)\e}\left|
1-\left(\frac{2(n-1)}{n+1}-\frac{(n-1)(n+3)\e}{2(n+1)}\right)x
\right|^2 \nonumber\\
\times\left(\frac{\pi }{\epsilon }+\pi  \left(x+\bar{x}-\log |x|^2\right)
-2 \pi  \epsilon  (x+\bar{x}) \log |x|^2\right)
\eea
The part corresponding to the intermediate field $\phi_{n,n+2}$ is simpler. For this
one we can restrict us with a less accurate expression for $I(x)$
\bea
I(x)\approx \frac{\pi }{\epsilon }-\pi \log |x|^2
\eea
and the integrand is simply
\bea
\frac{4 (n+3)^2 (1-(n+2) \epsilon )}{3 (n+1)^2}\left(
\frac{\pi }{\epsilon }-\pi \log |x|^2\nonumber
\right)\left|x\right|^{-2+2\e}
\eea
Performing integrations we get
\bea
&&\int_{D_{l,0}\backslash D_{l_0,0}} I(x) \langle \phi(x)\phi_{n,n+2}(0)\phi_{n,n+2}(1)\phi(\infty)
\rangle d^2x\nonumber\\
&&=\frac{8 \pi ^2 (n+2) (n+5) (n-1)^2}{3 n (n+1)^2 (n+3)^2 \epsilon ^2}
-\frac{\pi ^2 (n+2)}{3 l^2 n \epsilon }\nonumber\\
&&+\frac{4 \pi ^2 (n+2) (n-1) \left(\left(2 n^3+10 n^2+6 n-18\right)
\log (l)-n^3-9 n^2-23 n+1\right)}{3 n (n+1)^2 (n+3)^2 \epsilon }\nonumber\\
&&+\frac{8 \pi ^2 (n+3)^2}{3 (n+1)^2 \epsilon ^2}-
\frac{8 \pi ^2 (n+3)^2 (n+2-\log l)}{3 (n+1)^2 \epsilon }
\label{np2np20cont}
\eea
where the second and the third lines come from the family $[\phi_{n,n}]$
and the last line, from the $\phi_{n,n+2}$ field of the OPE (\ref{phiphinp2OPE}).\\
Also in this case the contribution coming from the region
$D_{l,\infty}$ is quite similar to the case discussed in Section \ref{bf}. We should simply take into account that
the contribution of the field $\phi$ appearing in the u-channel OPE is proportional to
\bea
C_{(1,3)(1,3)}^{(1,3)}C_{(1,3)(n,n+2)(n,n+2)}=
\frac{4 (n+3) (2-(n+5) \epsilon )}{3 (n+1)}+O(\e^2)
\eea
The result (c.f. eq. (\ref{infcontrib}))
is
\bea
&&\int_{D_{l,\infty}\backslash D_{l_0,\infty}} I(x) \langle \phi(x)\phi_{n,n+1}(0)\phi_{n,n+1}(1)\phi(\infty)
\rangle d^2x\nonumber\\
&&\approx -\frac{\pi^2}{\e l^2}+
\frac{4 \pi ^2 (n+3)}{3 (n+1)}\left(\frac{2}{\e^2}-
\frac{n+5-4 \log l}{\e}
\right)
\label{infcontnp2np2}
\eea
Finally, (\ref{np2np2Omegacont}) minus
(\ref{lensnp2np2}) plus twice (\ref{np2np20cont}) and plus (\ref{infcontnp2np2}) gives
\bea
&&\frac{8 \pi ^2 \left(3 n^4+33 n^3+121 n^2+143 n
+20\right)}{3 n (n+1) (n+3)^2 \epsilon ^2}\nonumber\\
&&-\frac{4 \pi ^2 (n+5) \left(5 n^4+45 n^3+143 n^2+151 n+8\right)}
{3 n (n+1) (n+3)^2 \epsilon }+O(\e^0)
\label{npm2C211}
\eea

\subsubsection{ Correlation function $\langle\phi_{n,n}(1)\phi_{n,n+2}(0)\rangle_\l$}
a) {\it{Contribution of the region}} $\Omega_{l,l_0}$\\
is given by the integral
\bea
\int_{\Omega_{l,l_0}} I(x)\langle\phi(x)\phi_{n,n+2}(0)\phi_{n,n}(1)\phi(\infty)\rangle d^2x
\label{Omegaintnnp2}
\eea
The large $p$ limit of the four-point function is very simple (see Appendix \ref{C})
\bea
\langle \phi(x)\phi_{n,n+2}(0)\phi_{n,n}(1)\phi(\infty)\rangle
=\frac{4}{3} \sqrt{\frac{n+2}{n}}\, |x|^{-2}+O(\e)
\label{eps0nnp2}
\eea
$I(x)$ can be replaced by
\bea
I(x)\approx -\frac{2\pi (n+1)\e}{n+5}\left|1+\frac{4 x}{(n+1) (1-x)} \right|^2
-\frac{2 \pi(n-3)\e}{n+1}\left|\frac{x}{1-x}\right|^2
\label{eps0nnp2I}
\eea
and the result of integration is
\bea
&&\int_{\Omega_{l,l_0}} I(x)\langle\phi(x)\phi_{n,n+2}(0)\phi_{n,n}(1)\phi(\infty)\rangle d^2x\nonumber\\
&&\approx \frac{ 16 \pi ^2 \epsilon }{3 (n+5)}\sqrt{\frac{n+2}{n}}
\left((3 n-5) \log (l)+(n+1) \log \left(2 l_0\right)\right)
\label{nnp2Omegacont}
\eea
b) {\it{Contribution of lens-like regions}} \\
We see from (\ref{eps0nnp2}), (\ref{eps0nnp2I}) that near $x\sim 1$ the integrand in eq. (\ref{Omegaintnnp2}) behaves as
\bea
-\frac{8 \pi  \e (n+1) \sqrt{\frac{n+2}{n}}}{3 (n+5)|x-1|^2}\nonumber
\eea
So according to the Appendix \ref{D} the contribution
of the lens-like regions near $x\sim 1$ which should be subtracted
from the r.h.s. of eq. (\ref{nnp2Omegacont}) is
\bea
-\frac{16 \pi^2  (n+1)\e}{3 (n+5)}\, \sqrt{\frac{n+2}{n}}
\, \log \left(\frac{l}{2 l_0}\right)
\label{lensnnp2}
\eea
c) {\it{Contributions of the regions}} $D_{l,0}$, $D_{l,1}$ {\it{and}} $D_{l,\infty}$\\
Let us compute the contributions of $D_{l,0}$ and $D_{l,1}$ separately.\\
i) $D_{l,0}$ {\it{contribution}}. \\
The relevant OPE has already appeared in (\ref{phiphinp2OPE}). Instead of (\ref{Lm1PhnPhnp2Ph})
we now need the analogous relation
\bea
\langle L_{-1}\phi_{n,n}(0)\phi_{n,n}(1)\phi(\infty)\rangle=
\left(2\D_{n,n}-\D\right)\langle \phi_{n,n}(0)\phi_{n,n}(1)\phi(\infty)\rangle
\label{Lm1PhnPhnPh}
\eea
The function $I(x)$ will be determined using the first equality in (\ref{Ieps}). For the calculation of the contribution of the field $\phi_{n,n+2}$ it is safe to replace the hypergeometric functions simply by $1$. Instead for the contribution of the family $[\phi_{n,n}]$ also the first order in $x$ and in $\bar{x}$ terms should be taken into account. Below we present expressions for the relevant combinations
of structure constants with required accuracy
\bea
&&C_{(1,3)(n,n+2)}^{(n,n)}C_{(1,3)(n,n)(n,n)}=\sqrt{\frac{n+2}{n}}\,
\frac{ \left(n^2-1\right)(2+5 \epsilon )\e^2}{48}+O(\e^4)\nonumber\\
&&C_{(1,3)(n,n+2)}^{(n,n+2)}C_{(1,3)(n,n+2)(n,n)}=
\sqrt{\frac{n+2}{n}}\,\frac{ (n+3) (2-(n+2) \epsilon)}{3 (n+1)}+O(\e^2)
\label{nnp2SC0}
\eea
For the final result of the $D_{l,0}$ contribution we get
\bea
&&\int_{D_{l,0}\backslash D_{l_0,0}} I(x) \langle \phi(x)\phi_{n,n+2}(0)\phi_{n,n}(1)\phi(\infty)
\rangle d^2x\nonumber\\
&&\approx -\frac{8 \pi ^2 (n-1) \sqrt{\frac{n+2}{n}}}
{3 (n+3) (n+5)}\left(1+ \epsilon
\left(n+\frac{5}{2}+(n+3) \log l\right)\right)\nonumber\\
&&-\frac{4 \pi ^2 (n+3) \sqrt{\frac{n+2}{n}}}{3 (n+5)}\left(1+
\epsilon  \left(
\frac{n}{2}+4+2 \log l\right)\right)
\label{nnp20cont}
\eea
where the two lines correspond to the  $[\phi_{n,n}]$
and $\phi_{n,n+2}$ contributions respectively.\\
ii) $D_{l,1}$ {\it{contribution}}. \\
The relevant OPE is
\bea
\label{phiphinOPE}
&&\phi(x)\phi_{n,n}(1)=|x-1|^{2(\D_{n,n+2}
-\D_{n,n}-\D)}C_{(1,3)(n,n)}^{(n,n+2)}
\phi_{n,n+2}(1)\qquad\qquad\qquad\\
&&+C_{(1,3)(n,n)}^{(n,n)}|x-1|^{-2\D}
\left(1+\frac{\D (x-1)}{2 \D_{n,n}}\,L_{-1} \right)
\left(1+\frac{\D (\bar{x}-1)}{2 \D_{n,n}}\,\bar{L}_{-1} \right)
\phi_{n,n}(1)
+\cdots\nonumber
\eea
The relation between the three point functions relevant for this case is
\bea
\langle L_{-1}\phi_{n,n}(1)\phi_{n,n+2}(0)\phi(\infty)\rangle=
\left(\D-\D_{n,n}-\D_{n,n+2}\right)\langle \phi_{n,n}(1)
\phi_{n,n+2}(1)\phi(\infty)\rangle \,\,\,\,
\label{Lm1Phn1Phnp2Ph}
\eea
Notice the flip of sign compared to (\ref{Lm1PhnPhnp2Ph}) due to
the rearrangement of the points $0$ and $1$.
For the function $I(x)$ the second equality in (\ref{Ieps}) should be used.
The hypergeometric functions should be expanded around $x=1$. When
calculating the contribution of $\phi_{n,n+2}$ it would suffice to keep
the constant term only while for the contribution of $[\phi_{n,n}]$ also
the terms linear in $x-1$ (or $\bar{x}-1$) should be taken into account.
During the calculation one encounters the same combinations
of the structure constants as in (\ref{nnp2SC0}).
Finally we get for the $D_{l,1}$ contribution a result identical to
that of $D_{l,0}$ given by (\ref{nm1np10cont}). Remember that
a similar phenomenon we have
encountered earlier in Section \ref{CFnm1np1}.\\
iii) $D_{l,\infty}$ {\it{contribution}}\\
Only the field $\phi$ appearing in u-channel OPE gives
a nonzero contribution. Since
\bea
C_{(1,3)(1,3)}^{(1,3)}C_{(1,3)(n,n)(n,n+2)}=
\sqrt{\frac{n+2}{n}}\,\,\, \frac{2(2-3 \e )}{3}
+O(\e^2),
\eea
and, from the third equality in (\ref{Ieps}),
\bea
I(x)\approx -\frac{4 \pi \e (n-3)}{n+5}\left(1+(n+5) \e \right)\,|x|^{-4 \epsilon }
\eea
we get
\bea
&&\int_{D_{l,\infty}\backslash D_{l_0,\infty}} I(x) \langle \phi(x)\phi_{n,n+2}(0)\phi_{n,n}(1)\phi(\infty)
\rangle d^2x\nonumber\\
&&\approx - \sqrt{\frac{n+2}{n}}\,\,\frac{8 \pi ^2 (n-3)
\left(2+\left(2n+7+4 \log l \right)\e\right)}{3 (n+5)}
\label{infcontnnp2}
\eea
It remains to combine all the contributions. The result is
\bea
- \sqrt{\frac{n+2}{n}}\,\,\frac{4 \pi ^2 (n-1)
\left(6n+22+\left(5 n^2+37n+64 \right)\e\right)}{3 (n+3)(n+5)}+O(\e^2)
\label{nnp2C212}
\eea

\subsubsection{ Correlation function $\langle\phi_{n,n+2}(1)\phi_{n,n-2}(0)\rangle_\l$}
a) {\it{Contribution of the region}} $\Omega_{l,l_0}$\\
\bea
\int_{\Omega_{l,l_0}} I(x)\langle\phi(x)\phi_{n,n-2}(0)\phi_{n,n+2}(1)\phi(\infty)\rangle d^2x
\label{Omegaintnp2nm2}
\eea
The large $p$ limit of the four-point function is (see Appendix \ref{C})
\bea
\langle \phi(x)\phi_{n,n-2}(0)\phi_{n,n+2}(1)\phi(\infty)\rangle
=\frac{\sqrt{n^2-4}}{3n|x(1-x)|^{4}}+O(\e)
\label{eps0np2nm2}
\eea
$I(x)$ can be replaced by
\bea
I(x)\approx -\frac{4\pi}{(n^2-4)\e}
\label{eps0np2nm2I}
\eea
and the result of the integration is
\bea
&&\int_{\Omega_{l,l_0}} I(x)\langle\phi(x)\phi_{n,n-2}(0)\phi_{n,n+2}(1)\phi(\infty)\rangle d^2x\nonumber\\
&&\approx \frac{\pi ^2}{6n\e\sqrt{n^2-4}}
\left(64 \log \left(2l\, l_0\right)+33
-\frac{8}{l^2}-\frac{4}{l_0^2}\right)
\label{np2nm2Omegacont}
\eea
b) {\it{Contribution of lens-like regions}} \\
It follows from (\ref{eps0np2nm2}) and (\ref{eps0np2nm2I})
that near $x\sim 1$ the integrand in eq. (\ref{Omegaintnp2nm2})
up to less singular terms behaves as
\bea
-\frac{4 \pi \left(1-2(x+\bar{x}-2)+(x-1)^2
+(\bar{x}-1)^2+4|x-1|^2\right) }{3n\e \sqrt{n^2-4}\,|x-1|^4 }
\nonumber
\eea
Consequently, from the Appendix \ref{D}, we see that the contribution of the
lens-like regions near $x\sim 1$ is
\bea
-\frac{4 \pi }{3n\e \sqrt{n^2-4} }
\left(\frac{\pi }{2 l_0^2}-\frac{\pi }{l^2}-\frac{\pi }{8}-2(\pi)+(2 \pi)+8 \pi  \log \left(\frac{l}{2 l_0}\right)\right)
\label{lensnp2nm2}
\eea
c) {\it{Contributions of the regions}} $D_{l,0}$, $D_{l,1}$ {\it{and}} $D_{l,\infty}$\\
i) $D_{l,0}$ {\it{contribution}} \\
The relevant OPE:
\bea
\label{phiphinm2OPE}
&&\phi(x)\phi_{n,n-2}(0)=(x\bar{x})^{\D_{n,n}-\D_{n,n-2}-\D}C_{(1,3)(n,n-2)}^{(n,n)}
\times\\
&&\left(1+\frac{\D+\D_{n,n}-\D_{n,n-2}}{2 \D_{n,n}}\,xL_{-1} \right)\left(1+\frac{\D+\D_{n,n}-\D_{n,n-2}}{2 \D_{n,n}}\,\bar{x}\bar{L}_{-1} \right)\phi_{n,n}(0)+\cdots\nonumber
\eea
The impact of $L_{-1}$ on the three-point function:
\bea
\langle L_{-1}\phi_{n,n}(0)\phi_{n,n+2}(1)\phi(\infty)\rangle=
\left(\D_{n,n}+\D_{n,n+2}-\D\right)\langle \phi_{n,n}(0)\phi_{n,n+2}(1)\phi(\infty)\rangle
\label{Lm1Phnp2Phnm2Ph}
\eea
In the first expression of (\ref{Ieps}) for $I(x)$,  the hypergeometric
functions should be expanded up to the linear order in
$x$ (or $\bar{x}$) terms.\\
The relevant combination of the structure constants:
\bea
C_{(1,3)(n,n-2)}^{(n,n)}C_{(1,3)(n,n)(n,n+2)}=\frac{\sqrt{n^2-4}}{3 n}+O(\e^2)\nonumber\\
\label{nm2np2SC0}
\eea
So, the final result for the $D_{l,0}$ contribution is
\bea
&&\int_{D_{l,0}\backslash D_{l_0,0}} I(x) \langle \phi(x)
\phi_{n,n-2}(0)\phi_{n,n+2}(1)\phi(\infty)
\rangle d^2x\nonumber\\
&&\approx \frac{16\pi ^2}{3 n \left(n^2-9\right) \sqrt{n^2-4}}
\left(\frac{10}{\e^2}+\frac{n^2-9}{4l^2\e}-\frac{n^2+1
+2(n^2-9)\log l}{\e}\right)\quad
\label{np2nm20cont}
\eea
ii) $D_{l,1}$ {\it{contribution}} \\
The relevant OPE:
\bea
\label{phiphinp2OPE1}
\phi(x)\phi_{n,n+2}(1)&=&|x-1|^{2(\D_{n,n}-\D_{n,n+2}-\D})C_{(1,3)(n,n+2)}^{(n,n)}\\
&\times & \left(1+\frac{\D+\D_{n,n}-\D_{n,n+2}}{2 \D_{n,n}}\,(x-1)L_{-1} \right)\nonumber\\
&\times & \left(1+\frac{\D+\D_{n,n}-\D_{n,n+2}}{2 \D_{n,n}}\,(\bar{x}-1)\bar{L}_{-1} \right)\phi_{n,n}(1)+\cdots\nonumber
\eea
The impact of $L_{-1}$ on the three-point function:
\bea
\langle L_{-1}\phi_{n,n}(1)\phi_{n,n-2}(0)\phi(\infty)\rangle=
\left(\D-\D_{n,n}-\D_{n,n-2}\right)\langle \phi_{n,n}(1)
\phi_{n,n-2}(1)\phi(\infty)\rangle \,\,\,\,
\label{Lm1Phn1Phnm2Ph}
\eea
The combination of structure constants required for this computation
coincides with that given by eq. (\ref{nm2np2SC0}).
The explicit calculation shows that in this case too, the $D_{l,1}$
contribution is identical to
that of $D_{l,0}$ given by (\ref{np2nm20cont}).\\
Note also that the contribution of $D_{l,\infty}$ is negligible.
Combining all the contributions for the case at hand we get
\bea
\frac{320\pi^2(1-\e)}{3\e^2 n(n^2-9)\sqrt{n^2-4}}+O(\e^2)
\label{nnp2C213}
\eea

\subsubsection{The matrix of anomalous dimensions}
The remaining two point functions $\langle\phi_{n,n-2}(1)\phi_{n,n-2}(0)\rangle_\l$ and $\langle\phi_{n,n}(1)\phi_{n,n-2}(0)\rangle_\l$ can be obtained from $\langle\phi_{n,n+2}(1)\phi_{n,n+2}(0)\rangle_\l$ and $\langle\phi_{n,n}(1)\phi_{n,n+2}(0)\rangle_\l$ replacing $n$ by $-n$.
So we have all necessary material to repeat the steps of Section \ref{MAD} and calculate
the matrix of anomalous dimensions for the fields
\bea
\phi_1\equiv \phi_{n,n+2}; \quad
\phi_2\equiv\left(2\D_{n,n}(2\D_{n,n}+1)\right)^{-1}
\partial\bar{\partial}\phi_{n,n};\quad
\phi_3\equiv\phi_{n,n-2}\nonumber
\eea
The two-point functions of these fields can be represented as in (\ref{npm12pf}),
but the indices now take the values $\a$,$\b=1,2,3$.
The replacement of the field $\phi_{n,n}$ by $\phi_{2}$ in a two point function,
at a given order $k$ of the perturbation theory,
results in an extra multiplier which is easy to calculate.
Here is the rule:
the coefficients $C^{(k)}_{\a,2}=C^{(k)}_{2,\a}$, for $\a\neq 2$, and $C^{(k)}_{2,2}$
should be endowed with the extra multipliers
\bea
\frac{\left(k\e-\D_{\a}-\D_{2}\right)^2}{2\D_{n,n}(2\D_{n,n}+1)}\nonumber
\eea
and
\bea
\left(\frac{\left(k\e-2\D_{2}\right)\left(k\e-2\D_{2}-1\right)}
{2\D_{n,n}(2\D_{n,n}+1)}\right)^2\nonumber
\eea
respectively. The numerators come from the derivatives and the
denominators from the normalization factor, present in the definition of the
field $\phi_2$.
The dimensions at the zero coupling $\l=0$ are
\bea
\D_1&=&\D_{n,n+2}=1-\frac{n+1}{2}\, \epsilon
+\frac{n^2-1}{16}\, \epsilon ^2+O(\e^3)\nonumber\\
\D_2&=&1+\D_{n,n}=1+\frac{n^2-1}{16}\,\epsilon ^2+O(\e^3)\nonumber\\
\D_3&=&\D_{n,n-2}=1+\frac{n-1}{2}\, \epsilon
+\frac{n^2-1}{16}\, \epsilon ^2+O(\e^3)
\eea

Computation of the first order coefficients as in previous cases
is quite easy and with desired accuracy we get
\bea
C^{(1)}_{1,1}&=&\frac{2 \pi  (n+3) (2-(n+2) \e )}{\sqrt{3} (n+1) \epsilon }
+O(\e)\nonumber\\
C^{(1)}_{1,2}&=&C^{(1)}_{2,1}=-\frac{8 \pi  \sqrt{\frac{n+2}{3n}}
\,(2-\epsilon )}{(n+1) (n+3) \epsilon }+O(\e)\nonumber\\
C^{(1)}_{1,3}&=&C^{(1)}_{3,1}=0;\,\,C^{(1)}_{2,2}=\frac{4 \pi
\left(4-(n^2 +1)\epsilon \right)}{\sqrt{3} (n^2-1)\epsilon }+O(\e)\nonumber\\
C^{(1)}_{2,3}&=&C^{(1)}_{3,2}=-\frac{8 \pi  \sqrt{\frac{n-2}{3n}}
(2-\epsilon )}{(n-3) (n-1) \epsilon }+O(\e)\nonumber\\
C^{(1)}_{3,3}&=&\frac{2 \pi  (n-3) (2+(n-2) \e )}
{\sqrt{3} (n-1) \epsilon }+O(\e)
\label{npm2C1}
\eea
From (\ref{npm2C211}), (\ref{nnp2C212}), (\ref{PhnPhnC2}),
(\ref{nnp2C213}) and the above presented
considerations, for the second order coefficients
$C^{(2)}_{\a,\b}=C^{(2)}_{\b,\a}$ we find

\bea
C^{(2)}_{1,1}&=&\frac{8 \pi ^2 \left(3 n^4+33 n^3+121 n^2
+143 n+20\right)}{3 n (n+1) (n+3)^2 \epsilon ^2}\nonumber\\
&-&\frac{4 \pi ^2 (n+5) \left(5 n^4+45 n^3+143 n^2+151 n+8\right)}
{3 n (n+1) (n+3)^2 \epsilon }+O(\e^0)\nonumber\\
C^{(2)}_{1,2}&=&-\frac{64 \pi ^2 \sqrt{\frac{n+2}{n}} (3 n+11)}{3 (n+1) (n+3) (n+5)\,\, \epsilon ^2}+\frac{32 \pi ^2 \sqrt{\frac{n+2}{n}}
\left(n^2+18 n+57\right)}{3 (n+1) (n+3) (n+5) \epsilon }
+O(\e^0)\nonumber\\
C^{(2)}_{1,3}&=&\frac{320 \pi ^2}{3 n(n^2-9) \sqrt{n^2-4}\,\, \epsilon ^2}
-\frac{320 \pi ^2}{3n (n^2-9) \sqrt{n^2-4}\,\, \epsilon }+O(\e^0)\nonumber\\
C^{(2)}_{2,2}&=&\frac{128 \pi ^2}{3 \left(n^2-1\right) \epsilon ^2}
-\frac{16\pi^2 \left( n^2+19 \right)}{3 \left(n^2-1\right) \epsilon }
+O(\e^0)\nonumber\\
C^{(2)}_{2,3}&=&-\frac{64 \pi ^2 \sqrt{\frac{n-2}{n}}
(3 n-11)}{3 (n-1) (n-3) (n-5) \epsilon ^2}
-\frac{32 \pi ^2 \sqrt{\frac{n-2}{n}}
\left(n^2-18 n+57\right)}{3 (n-1) (n-3) (n-5) \epsilon }
+O(\e^0)\nonumber\\
C^{(2)}_{3,3}&=&\frac{8 \pi ^2 \left(3 n^4-33 n^3+121 n^2
-143 n+20\right)}{3 n (n-1) (n-3)^2 \epsilon ^2}\nonumber\\
&+&\frac{4 \pi ^2 (n-5) \left(5 n^4-45 n^3+143 n^2-151 n+8\right)}
{3 n (n-1) (n-3)^2 \epsilon }+O(\e^0)
\label{npm2C2}
\eea
With this input we can repeat the procedure of the
Section \ref{MAD} and compute the matrix of anomalous dimensions.
Here is the final result:
\bea
\G_{1,1}&\approx &\D_1+\frac{\pi g (n+3)(2-(n+2)\e)}{\sqrt{3}\, (n+1)}
+\frac{8 \pi^2g^2 (n+2)}{3 (n+1)}\nonumber\\
\G_{1,2}&=&\G_{2,1}\approx \frac{\pi g (n-1) \sqrt{\frac{n+2}{3n}}\,\,
(2-\e)}{n+1}+\frac{4 \pi ^2 g^2(n-1)
\sqrt{\frac{n+2}{n}}}{3 (n+1)}\nonumber\\
\G_{1,3}&=&\G_{3,1}\approx 0\nonumber\\
\G_{2,2}&\approx &\D_2+\frac{2 \pi g(4-(n^2+1)\e)}{\sqrt{3}\, (n^2-1)}
+\frac{4 \pi ^2 g^2\left(n^2+3\right)}{3 (n^2-1)}\nonumber\\
\G_{2,3}&=&\G_{3,2}\approx \frac{\pi g (n+1) \sqrt{\frac{n-2}{3n}}\,\,
(2-\e)}{ n-1}+\frac{4 \pi ^2 g^2(n+1)
\sqrt{\frac{n-2}{n}}}{3 (n-1)}\nonumber\\
\G_{3,3}&\approx &\D_3+\frac{\pi g (n-3)(2+(n-2)\e)}{\sqrt{3}\, (n-1)}
+\frac{8 \pi^2g^2 (n-2)}{3 (n-1)}
\eea
Again we see that all matrix elements are regular at $\e=0$.
All double and single poles in $\e$  disappeared.
At the fixed point $g=g^*$ (see (\ref{gfp}))
\bea
\G_{1,1}^{(g^*)}&=&1-\frac{\left(n^2-5\right) \epsilon }{2 (n+1)}+\frac{\left(n^3-7 n^2-n+39\right) \epsilon ^2}{16 (n+1)}+O(\e^3)\nonumber\\
\G_{1,2}^{(g^*)}&=&\G_{2,1}^{(g^*)}=
\frac{(n-1) \sqrt{\frac{n+2}{n}}\,\,
 \epsilon  (\epsilon +1)}{n+1}+O(\e^3)\nonumber\\
\G_{1,3}^{(g^*)}&=&\G_{3,1}^{(g^*)}=O(\e^3)\nonumber\\
\G_{2,3}^{(g^*)}&=&\G_{3,2}^{(g^*)}=
\frac{(n+1) \sqrt{\frac{n-2}{n}}\,\,
 \epsilon  (\epsilon +1)}{n-1}+O(\e^3)\nonumber\\
\G_{3,3}^{(g^*)}&=&1+\frac{\left(n^2-5\right)
\epsilon }{2 (n-1)}+\frac{\left(n^3+7 n^2-n-39\right)
\epsilon ^2}{16 (n-1)}+O(\e^3)
\label{Gammanpm2}
\eea
Here are the eigenvalues of this matrix
\bea
\D_1^{(g^*)}&=&1+\frac{(n+1)\e}{2}+\frac{(n+1)(n+7) \e ^2}{16}+O(\e^3)
\nonumber\\
\D_2^{(g^*)}&=&1+\frac{ (n^2-1)\e^2}{16}+O(\e^3)\nonumber\\
\D_3^{(g^*)}&=&1-\frac{(n-1)\e}{2}+\frac{(n-1)(n-7) \e^2}{16}+O(\e^3)
\eea
which, up to $O(\e^3)$ terms coincide with the dimensions $\D_{n+2,n}^{(p-1)}$,
$1+\D_{n,n}^{(p-1)}$ and $\D_{n-2,n}^{(p-1)}$ of the IR CFT $M_{p-1}$.
It is easy to find the orthogonal matrix which diagonalizes  the matrix
of anomalous dimensions (\ref{Gammanpm2})
and to establish the explicit map
\bea
\phi_{n+2,n}^{(p-1)}&=&\frac{2}{n^2+n}\,\,\phi_{1}^{(g^*)}+
\frac{2 \sqrt{\frac{n+2}{n}}}{n+1}\,\,\phi_{2}^{(g^*)}+
\frac{\sqrt{n^2-4}}{n}\,\,\phi_{3}^{(g^*)}\nonumber\\
\phi_2^{(p-1)}&=&-\frac{2 \sqrt{\frac{n+2}{n}}}{n+1}\,\,\phi_{1}^{(g^*)}+
\frac{5-n^2}{n^2-1}\,\,\phi_{2}^{(g^*)}+
\frac{2 \sqrt{\frac{n-2}{n}}}{n-1}\,\,\phi_{3}^{(g^*)}\nonumber\\
\phi_{n-2,n}^{(p-1)}&=&\frac{\sqrt{n^2-4}}{n}\,\,\phi_{1}^{(g^*)}
-\frac{2 \sqrt{\frac{n-2}{n}}}{n-1}\,\,\phi_{2}^{(g^*)}+
\frac{2}{(n-1) n}\,\,\phi_{3}^{(g^*)}
\label{npm2map}
\eea
where
\bea
\phi_2^{(p-1)}\equiv \left(2\D_{n,n}^{(p-1)}(2\D_{n,n}^{(p-1)}+1)\right)^{-1}
\partial\bar{\partial}\phi_{n,n}^{(p-1)}
\eea
In this case too we see that the coefficients in (\ref{npm2map}) do not receive
$\e$ or $\e^2$ corrections.
\section*{ Acknowledgements}
Its a pleasure to thank R.~Flume for interesting discussions.
This work was partly supported by European Commission FP7 Programme Marie Curie
IIF Return Phase Grant Agreement 908571, by Volkswagen foundation of Germany, by
a grant of the Armenian State Council of Science and by Armenian-Russian grant
"Common projects in Fundamental Scientific Research"-2013.
\section*{ Note added}
After this paper was published Stefan Fredenhagen
pointed out that the next to leading
order computations carried out in this paper control only
$1/p$ corrections to the mixing matrix. But such corrections
are absent also in Gaiotto's RG domain wall approach.
Thus contrary to what was claimed in the main text,
our results are in complete agreement with the RG domain
wall approach and there is no need to involve
another regularization scheme.
I am grateful to Stefan Fredenhagen for this clarification.  
\begin{appendix}
\section{Minimal models}
\label{A}
For the readers convenience we present here few facts about unitary
series of the minimal models \cite{BPZ:1984,FQS:1984} denoted by
$M_p$, $p=3,4,\ldots$ . The central charge is given by
\bea
c_p=1-\frac{6}{p(p+1)}
\eea
This theory contains finitely many spinless\footnote{We consider
here the so called diagonal series only.} primary fields denoted by $\phi_{n,m}$
with conformal
dimensions (the famous Kac spectrum \cite{Kac})
\bea
\D_{n,m}=\frac{(n-m)^2}{4}+\frac{n^2-1}{4p}-\frac{m^2-1}{4(p+1)}
\label{DimGen}
\eea
where $n\in \{1,2,\ldots, p-1\}$, $m\in \{1,2,\ldots ,p\}$.
There is an identification $\phi_{p-n,p+1-m}\equiv \phi_{n,m}$ so
that the number of primary fields is equal to $p(p-1)/2$. The field
$\phi_{1,1}$ with dimension $0$ is the identity operator. The operator product
expansions satisfy the fusion rules
\bea
\phi_{n_1,m_1}\phi_{n_2,m_3}\in \mathop{\bigoplus_{n_3=|n_1-n_2|
+1}^{n_1+n_2-1}}_{n_1+n_2-n_3\in 2\mathbb{Z}+1}\quad\mathop{\bigoplus_{m_3=|m_1-m_2|
+1}^{m_1+m_2-1}}_{m_1+m_2-m_3\in 2\mathbb{Z}+1}[\phi_{n_3,m_3}]
\eea
The main subject of this paper is the minimal model $M_p$
perturbed by the relevant field $\phi_{1,3}$. It's dimension
\bea
\D_{1,3}=1-\frac{2}{p+1}\equiv 1-\e <1
\eea
For large $p$ this field becomes nearly marginal which is the main
reason why in this region the non-trivial RG behavior can be investigated
by means of the perturbation theory.
The structure constants of the OPE have been computed in \cite{DF:1985SC}. A slightly
more compact expression which we present below is taken from \cite{Pog:1989SC}
\bea
\label{StrConstGen}
&&C_{(n_1,m_1)(n_2,m_2)(n_3,m_3)}=\rho^{4st+2t-2s-1}\\
&&\times \sqrt{\frac{\gamma(\rho-1) \gamma(m_1-n_1\rho^{-1})
\gamma(m_2-n_2\rho^{-1})\gamma(-m_3+n_3\rho^{-1})}
{\gamma(1-\rho^{-1}) \gamma(-n_1+m_1\rho)
\gamma(-n_2+m_2\rho)\gamma(n_3-m_3\rho)}}\nonumber\\
&&\times\prod_{i=1}^s\prod_{j=1}^t\left((i-j\rho)(i+n_3-(j+m_3)\rho)(i-n_1-(j-m_1)\rho)
(i-n_2-(j-m_2)\rho)\right)^{-2}\nonumber\\
&&\times\prod_{i=1}^s\gamma(i\rho^{-1})\gamma(-m_3+(i+n_3)\rho^{-1})
\gamma(m_1+(i-n_1)\rho^{-1})\gamma(m_2+(i-n_2)\rho^{-1})\nonumber\\
&&\times\prod_{j=1}^t\gamma(j\rho)\gamma(-n_3+(j+m_3)\rho)
\gamma(n_1+(j-m_1)\rho)\gamma(n_2+(j-m_2)\rho)\nonumber
\eea
where
\bea
\gamma(x)\equiv\frac{\Gamma(x)}{\Gamma(1-x)};\quad
s=\frac{n_1+n_2-n_3-1}{2}; \qquad t=\frac{m_1+m_2-m_3-1}{2}\nonumber
\eea
\section{Computation of $I(x)$}
\label{B}
One way to get the result (\ref{Iabc}) for the integral (\ref{I(x)}) is to
notice that $I(x)$ satisfies the hypergeometric differential equation independently
with respect to the both variables $x$ and $\bar{x}$. The starting point is the identity
\bea
&&\left[x(1-x)\partial_x^2+(1-a-c+(a+b+2c-2)x)\partial_x\right.\\
&&\left.+c(1-a-b-c)\right]
\left(y^{a-1}(1-y)^{b-1}(y-x)^{c}\right)
=\partial_y\left(c\,y^{a}(1-y)^{b}(y-x)^{c-1}\right)\nonumber
\eea
which shows that as a function of the variable $x$, $I(x)$ is a linear combination
of the hypergeometric functions
\bea
F(1-a-b-c,-c,1-a-c,x)\nonumber
\eea
and
\bea
x^{a+c}F(a,1-b,1+a+c,x)\nonumber
\eea
The same conclusion is true also for the conjugate variable $\bar{x}$. The condition
that the function $I(x)$ is single valued around the points $x=0$ a $x=1$ fixes a specific combination of holomorphic and anti-holomorphic
parts up to a constant
which in its turn can be easily evaluated considering the special case $x=0$. The
final result is presented in eq. (\ref{Iabc}).

\section{Four-point functions at large $p$ limit}
\label{C}
Since the structure constants of OPE for the minimal models are
known (see \ref{StrConstGen}), to construct the correlation
functions it remains to calculate related conformal blocks. According
to AGT relation \cite{AGT:2009} this conformal blocks in a simple fashion are related
to the instanton part of the Nekrasov partition function of $N=2$ SYM
theory with the gauge group $SU(2)$ and with four fundamental hypermultiplets.
In the large $p$ limit the minimal
models approach to a free theory (the central charge $c\approx 1$), so it
is not surprising that in this limit conformal blocs of degenerated
primary fields become very simple and can be expressed in terms of rational
(and also logarithmic in the cases when the leading corrections in $1/p$ is
required to be taken into account) functions of the the cross ratio of
the coordinates. It is straightforward to compute Nekrasov partition
\cite{Nekrasov:2002} function
up to desired order in instanton expansion using combinatorial
formula found in \cite{FP:2002} and extended to the case with extra hypermultiplets
in \cite{BFMT:2002}. Computing the first few coefficients of the instanton expansion
(for more confidence we made calculations up to $6$th order ),
adjusting appropriately the parameters in order to get the required conformal block
and finally taking the large $p$ limit one can easily
guess the exact dependence of the conformal block on the cross ratio of the insertion
points (which is the same as the instanton counting parameter, from the gauge
theory point of view). In this way we got expressions\footnote{Some particular conformal blocs in large $p$ limit have been computed
earlier in \cite{Konechny:2012} using more traditional approach. }
for the correlation functions
\bea
&&\langle \phi _{1,3}(x) \phi _{1,3}(0) \phi _{n,n}(1) \phi _{n,n}(\infty)\rangle=
|x|^{4 \epsilon -4}\nonumber\\
&&+\frac{\left(n^2-1\right) \epsilon ^2}{12 |x|^4}
\left(\frac{x^2}{2 (1-x)}+\frac{\bar{x}^2}{2 (1-\bar{x})}+\log ^2|1-x|^2\right)+O(\e^3)\nonumber\\
&&\langle \phi _{1,3}(x) \phi _{1,3}(0)
\phi _{n,n+1}(1) \phi _{n,n+1}(\infty)\rangle=
\left|\frac{1-x+\frac{x^2}{2}}{x^2(1-x)}\right|^2+\frac{2(n+2)}{3n}
\left|\frac{1-\frac{x}{2}}{x(1-x)}\right|^2+O(\e)\nonumber\\
&&\langle \phi _{1,3}(x) \phi _{1,3}(0) \phi _{n,n-1}(1)
\phi _{n,n+1}(\infty)\rangle=
\frac{4 \sqrt{n^2-1}}{3n} \left|
\frac{1-\frac{x}{2}}{x(1-x)}\right|^2+O(\e)\nonumber\\
&&\langle\phi _{1,3}(x) \phi _{1,3}(0)
\phi _{n,n+2}(1) \phi _{n,n+2}(\infty)\rangle\nonumber\\
&&=\left|\frac{1-2x+3x^2-2x^3+\frac{x^4}{3}}{x^2(1-x)^2}\right|^2
+\frac{8 (n+3)}{3 (n+1)}\left|\frac{1-\frac{3x}{2}
+ x^2 -\frac{ x^3}{4}}{x(1-x)^2}\right|^2\nonumber\\
&&+\frac{(n+3) (n+4)}{18 n (n+1)}\left|\frac{x}{1-x}\right|^4+O(\e)\nonumber\\
&&\langle\phi _{1,3}(x) \phi _{1,3}(0) \phi _{n,n}(1)
\phi _{n,n+2}(\infty)\rangle=\frac{4}{3}
\sqrt{\frac{n+2}{n}}|x|^{-2}+O(\e)\nonumber\\
&&\langle\phi _{1,3}(x) \phi _{1,3}(0) \phi _{n,n+2}(1)
\phi _{n,n-2}(\infty)\rangle= \frac{\sqrt{n^2-4}}{3n}
\left|\frac{x}{1-x}\right|^{4}+O(\e)
\label{4pfinfp}
\eea
As a nontrivial check, we have tested
the crossing invariance of all these correlation functions. Of course
the interested reader can get convinced in correctness of our expressions
also by examining the third order differential equation satisfied by any conformal
bloc which includes the degenerated field $\phi_{1,3}$ \cite{BPZ:1984}.

Performing the conformal map $x\rightarrow 1/x$ with the help of the eq. (\ref{Ginf})
we get the correlation functions  (\ref{G0eps}), (\ref{Gneps}), (\ref{eps0np1np1}), (\ref{Gneps}), (\ref{eps0nm1np1}), (\ref{eps0np2np2}),
(\ref{eps0nnp2}), (\ref{eps0np2nm2}) used in the main text.

\section{Integrations over lens-like regions}
\label{D}
\begin{figure}[htb]
\includegraphics[scale=0.3]{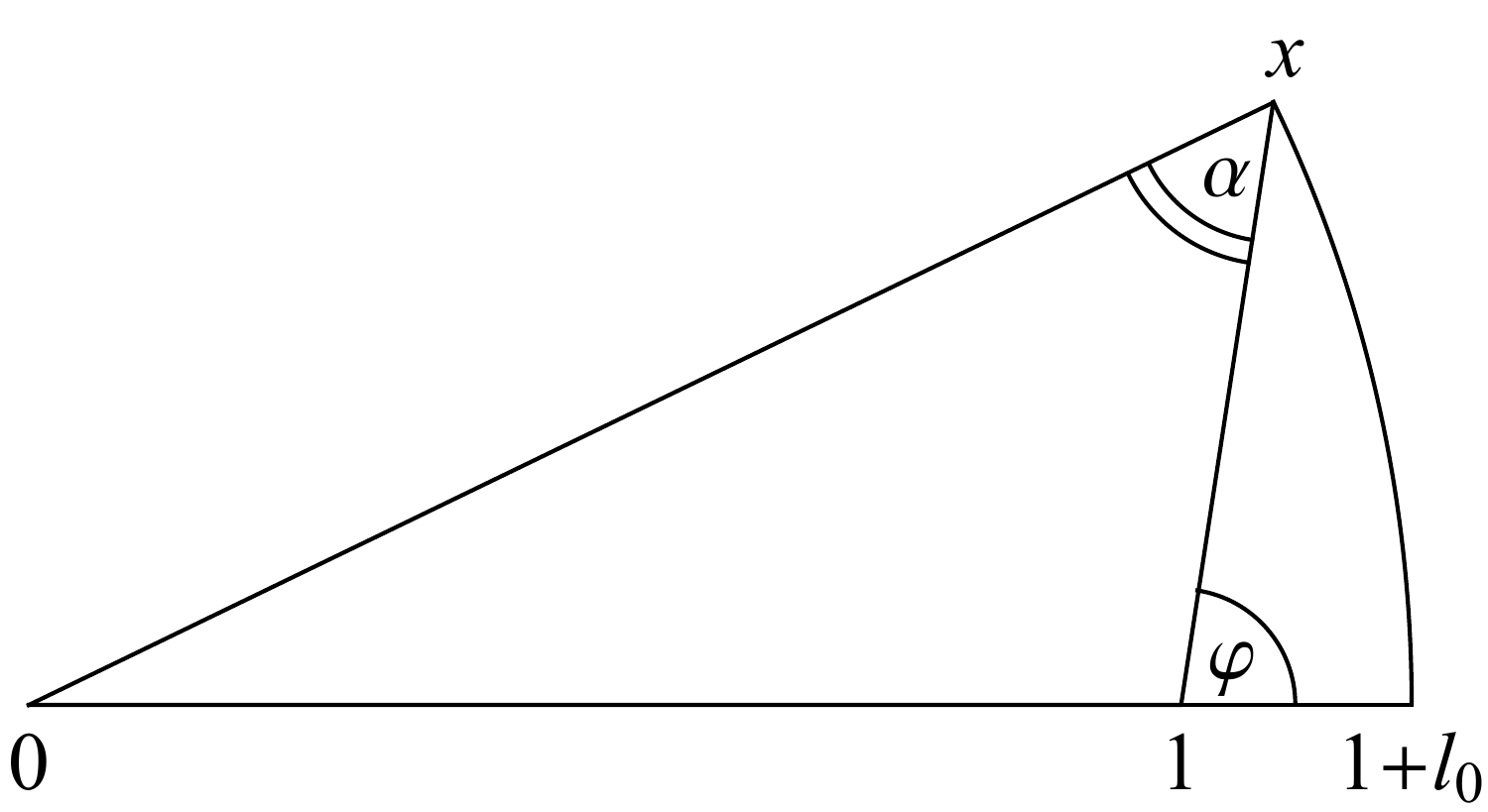}
\centering
\caption{The angles $\phi$ and $\alpha$; $|x|=1+l_0$}
\label{Fig2}
\end{figure}

Here we compute the contributions of the lens-like regions
\bea
D_L=D_{1-l_0,0}\cap D_{l,1}\nonumber\\
D_R=D_{1+l_0,0}\cap D_{l,1}\nonumber\\
\eea
(see Fig.\ref{Fig1}
and also the discussion coming after the eq. (\ref{Omegaint})). Using
Green's theorem the integrals over lens-like regions can be easily transformed
to the contour integrals over their boundaries. The integrals over the arcs
which belong to the circle $|x-1|=l$ are trivial. Instead, the integrals along
remaining parts of the boundary which lay on $|x|=1+l_0$ (for the right
lens-like region $D_R$) or on $|x|=1-l_0$ (for the left lens-like region $D_L$) seem
more complicated, but fortunately these contour integrals too (with
an exception to be considered later) admit exact treatment. Below we give
the details on the integration along the arc $|x|=1+l_0$. The formulae for
the other arc $|x|=1-l_0$ can be found by a simple replacement
$l_0\leftrightarrow -l_0$. During the calculations we heavily employ the
formulae (we use the
notation $r\equiv |x-1|$ and the angles $\phi$, $\a$ are depicted in
Fig.\ref{Fig2})
\bea
&&(1+l_0)\sin(\a)=\sin(\varphi); \quad r=(1+l_0)\cos(\a)-\cos(\varphi)\\
&&r^{-1}=\frac{(1+l_0)\cos(\a)+\cos(\varphi)}{l_0(l_0+2)}
\eea

First note that for the arbitrary region ${\cal{D}}$
\bea
&&\int_{{\cal{D}}}\frac{d^2x}{|x-1|^4}=
\int_{\partial{\cal{D}}}\frac{drd\varphi}{r^3}=-\frac{1}{2}
\int_{\partial{\cal{D}}}\frac{d\varphi}{r^2}\nonumber\\
&&\int_{{\cal{D}}}d^2x\left(\frac{1}{(x-1)^2}+\frac{1}{(\bar{x}-1)^2}\right)
=-\int_{\partial{\cal{D}}}\frac{\sin(2\varphi)dr}{r}\nonumber\\
&&\int_{{\cal{D}}}\frac{d^2x}{|x-1|^2}\left(\frac{1}{x-1}+\frac{1}{\bar{x}-1}\right)
=-\int_{\partial{\cal{D}}}\frac{2d\sin(\varphi)}{r}\nonumber\\
\eea
When restricted on the circle $|x|=1+l_0$ the one forms appearing
on the r.h.s. can be represented as total derivatives:
\bea
&&\frac{d\varphi}{r^2}=\frac{(1+l_0)^2\cos^2(\a)+\cos^2(\varphi)
+2(1+l_0)\cos(\a)\cos(\varphi)}{l_0^2(1+l_0)^2}\,\,d\varphi\nonumber\\
&&=\frac{(1+l_0)^2d(\a+\varphi)+(1+l_0)d\sin(\a+\varphi)}
{l_0^2(2+l_0)^2}\nonumber\\
&&\frac{\sin(2\varphi)dr}{r}=\frac{(1+l_0)^2}{2}d(2\a-\sin(2\a))\nonumber\\
&&\frac{2d\sin(\varphi)}{r}=d\left(\frac{(1+l_0)^2\a+\varphi
+(1+l_0)\sin(\a+\varphi)}{l_0(l_0+2)}\right)
\eea
With these formulae at hand it is easy to evaluate the integrals over lens-like
regions in the limit $l_0/l\rightarrow 0$
and $l\rightarrow0$
\bea
&&\int_{D_L\cup D_R}\frac{d^2x}{|x-1|^4}\approx \left(-\frac{\pi }{l^2}-\frac{\pi }{8}\right)+\frac{\pi }{2 l_0^2}\\
&&\int_{D_L\cup D_R}d^2x\left(\frac{1}{(x-1)^2}+\frac{1}{(\bar{x}-1)^2}\right)
\approx 2\pi\nonumber\\
&&\int_{D_L\cup D_R}\frac{d^2x}{|x-1|^2}\left(\frac{1}{x-1}+\frac{1}{\bar{x}-1}\right)
\approx \pi\nonumber\\
\eea
We need also the integral
\bea
\int_{{\cal{D}}}\frac{d^2x}{|x-1|^2}=\int_{\partial{\cal{D}}} \log(r)d\varphi
\eea
Unlike the previous cases this integral can not be evaluated exactly
in terms of elementary functions. Nevertheless it is not difficult to
show that up to terms vanishing in the limit $l_0/l\rightarrow 0$
and $l\rightarrow0$ it is equal to
\bea
\int_{D_L\cup D_R}\frac{d^2x}{|x-1|^2}\approx 2\pi \log\left(\frac{l}{2l_0^2}\right)
\label{LensLog}
\eea


\end{appendix}
\bibliographystyle{JHEP}

\begin{thebibliography}{99}

\bibitem{Zamolodchikov:1987}
A.~Zamolodchikov, {\it {Renormalization Group and Perturbation Theory Near
  Fixed Points in Two-Dimensional Field Theory}},  {\em Sov.J.Nucl.Phys.} {\bf
  46} (1987) 1090.

\bibitem{Poghossian:1988}
R.~Poghossian, {\it{Study of the Vicinities of Superconformal Fixed Points
in Two-dimensional Field Theory}}, {\em Sov.J.Nucl.Phys. } {\bf 48}(1988) 763.

\bibitem{Gaiotto:2012RGDW}
D.~Gaiotto, {\it {Domain Walls for Two-Dimensional Renormalization Group Flows}},  [\href{http://xxx.lanl.gov/abs/1203.1052}{{\tt arXiv:1203.1052 [hep-th]}}].

\bibitem{Zamolodchikov:1986}
A.~B.~Zamolodchikov, {\it{"Irreversibility" of the flux of the renormalization
group in a 2D field theory}}, {\em JETP Lett.}{\bf 43}(12), 730-732 (1986).

\bibitem{BPZ:1984}
A.~Belavin, A.~Polyakov and A.~Zamolodchikov, {\it {Infinite conformal symmetry in two-dimensional quantum field theory
}},  {\em Nucl.Phys.} {\bf B241} (1984).
  333--380.
\bibitem{Kac}
V.~G.~Kac, {\it{Highest weight representations of infinite dimensional
Lie algebras}}, Proc. Internat. Congress Mathematicians (Helsinki, 1978).

\bibitem{BatErd1}
A.~Erdelyi et al.,{\it {Higher transcendental functions}}, vol. 1,
(McGraw-Hill Book Co., Inc., New York, N.Y., 1953).

\bibitem{FQS:1984}
D.~Friedan, Z.~Qiu and S.~Shenker, {\it {
Conformal Invariance, Unitarity, and Critical Exponents in Two Dimensions
}},  {\em Phys.Rev.Lett.} {\bf v. 52} (1984)
  1575--1578.
\bibitem{DF:1985SC}
Vl.~Dotsenko, V.~Fateev, {\it {Operator algebra of two-dimensional conformal theories with central charge $C\le 1$
}},  {\em Phys.Lett.} {\bf B154} (1985)
  291--295.

\bibitem{Pog:1989SC}
R.~G.~Pogosian, {\it{Fields with spin in the minimal models $M(p)$ ($c<1$) of two-dimensional conformal field theory}},  {\em preprint} YERPHI-1198-75-89. [\href{http://ccdb5fs.kek.jp/cgi-bin/img_index?199008434}{{KEK library link}}]

\bibitem{AGT:2009}
L.~Alday, D.~Gaiotto, and Y.~Tachikawa, {\it{Liouville Correlation
Functions from Four-dimensional Gauge Theories}}, {\em{Lett. Math. Phys.}} {\bf 91}
(2010) 167--197,
[\href{http://xxx.lanl.gov/abs/0906.3219}{{\tt arXiv:0906.3219 [hep-th]}}].

\bibitem{Nekrasov:2002}
N.~Nekrasov, {\it{Seiberg-Witten prepotential from instanton counting}},
  in {\em Adv.Theor.Math.Phys.}{\bf 7} (2004): 831-864,
  [\href{http://xxx.lanl.gov/abs/hep-th/0206161}{{\tt hep-th/0206161}}].

\bibitem{FP:2002}
R.~Flume and R.~Poghossian, {\it{An algorithm for the microscopic evaluation
of the coefficients of the Seiberg-Witten prepotential}},
{\em{Int.J.Mod.Phys.}} {\bf A18} (2003) 2541,
[\href{http://xxx.lanl.gov/abs/hep-th/0208176}{{\tt hep-th/0208176}}].

\bibitem{BFMT:2002}
U.~Bruzzo, F.~Fucito, J.~F. Morales, and A.~Tanzini, {\it{Multi-instanton
  calculus and equivariant cohomology}}, {{\em JHEP}} 0305:054,2003,
  [\href{http://xxx.lanl.gov/abs/hep-th/hep-th/0211108}{{\tt hep-th/0211108}}].

\bibitem{Konechny:2012}
  A.~Konechny, {\it{Renormalization group defects for boundary flows}},
  {{\em J.Phys.A}} {\bf 46}, 145401, (2013),
  [\href{http://xxx.lanl.gov/abs/1211.3665}{{\tt arXiv:1211.3665 [hep-th]}}].
\end{thebibliography}
\providecommand{\href}[2]{#2}
\begingroup\raggedright

\endgroup

\end{document}